\documentclass{article}

\usepackage[utf8]{inputenc}
\usepackage{graphicx}
\usepackage{natbib}
\usepackage{hyperref}
\usepackage{xr}

\usepackage[english]{babel}
\usepackage{float}
\usepackage{amsmath}
\usepackage{fullpage}
\usepackage{caption}
\usepackage{ifthen}
\usepackage[format=hang]{subfig}
\usepackage{threeparttablex}
\usepackage{tabularx}
\usepackage{longtable}
\usepackage{booktabs}
\usepackage{geometry}
\usepackage{lscape}
\usepackage{afterpage}
\usepackage{xcolor}
\usepackage{placeins}
\usepackage{multicol}
\usepackage{supertabular}
\usepackage{amssymb}

\setcitestyle{authoryear,open={(},close={)}} 
\setcitestyle{citesep={;}}

\definecolor{magenta_custom}{rgb}{1, 0, 1}
\definecolor{green_custom}{rgb}{0, 0.7, 0}

\graphicspath{ {images/} }

\begin{document}

\title{Characterising injection signatures in Jupiter’s ultraviolet aurora using \textit{Juno} observations}
\author{
    L. A. Head \thanks{Corresponding author: LA.Head@uliege.be (email address)},
    D. Grodent,
    B. Bonfond,
    A. Moirano,
    G. Sicorello,\\
    J. Vinesse,
    A. Mouton,
    M. Dumont,
    T. K. Greathouse,
    V. Hue,\\
    A. Sulaiman,
    B. H. Mauk,
    Z. H. Yao,
    R. L. Guo,
    J. Y. Zhao
}

\date{\today}

\abstract{
    Discrete features in Jupiter's ultraviolet aurora have been interpreted as signatures of plasma injections in the middle magnetosphere.
    There exists some ambiguity whether magnetodisc scattering or high-latitude Alfv\'{e}nic acceleration best describes the observed properties of these injection signatures, and also to what extent arcs in the outer emission are related to injections.
    Many injection signatures are the result of the evolution of dawn storms; there is, however, limited evidence that non-dawn-storm injection signatures are sometimes present in the aurora.
    We use automatic detection of these discrete features, alongside data from \textit{Juno}-UVS and in-situ measurements by other \textit{Juno} instruments, to show that scattering likely accounts for most of the electron precipitation associated with injection signatures.
    Additionally, there is evidence that injection signatures can be classified into two types: dawn-storm and non-dawn-storm.
    Arc-like features in the outer emission show very similar properties to traditional blob-like injection signatures and may consist of sequences of injection signatures that have broadened into an arc via energy-dependent electron drift.
}

\maketitle

\section{Introduction}
\label{sec:introduction}

In the equatorward region or outer emission of Jupiter's ultraviolet (UV) aurora, discrete or patch-like features are frequently observed that can be linked to plasma injections in Jupiter's magnetosphere \citep{dumont+:2014}.
These plasma injections were first detected by the \textit{Galileo} probe \citep{mauk+:1997} and are expected to arise from interchange instability, where hot plasma from the distant magnetosphere moves inward to compensate for the outward flow of new plasma from Io \citep{paranicas+:1991}.  
Plasma injections frequently, but not always \citep{haggerty+:2019}, give rise to injection signatures in Jupiter's UV aurora.
These signatures typically have an amorphous or blob-like morphology and can, on occasion, emit more power than the main emission \citep{palmaerts+:2023}.  
The prevailing understanding suggests that anisotropic electron distributions induced by interchange events (injections) favour the production of plasma waves in the equatorial plane \citep{daly+:2023}.
These waves impose isotropic pitch-angle scattering on the magnetospheric electrons, filling the loss cone, leading to electron precipitation in the ionosphere and hence auroral emission \citep{li+:2021} which has been directly observed in a young plasma injection by \textit{Juno} \citep{menietti+:2021}.
However, an alternative high-latitude mechanism has been suggested to explain certain characteristics of injection signatures \citep{gray+:2017,dumont:2023} and could provide sufficient energy flux to power the equatorward aurora \citep{gershman+:2019}.
In this second scenario, changes in the magnetic-field topology induced by an injection may launch Alfv\'{e}n waves toward the ionosphere, where they accelerate electrons at high latitude and provoke auroral emission \citep{gray+:2017}.
It is not yet entirely clear to what extent these two mechanisms contribute to injection signatures; the diffuse aurora, in which injection signatures are sometimes embedded, appears, to some extent, to result from high-latitude Alfv\'{e}nic acceleration, at least at low electron energies \citep{sulaiman+:2022,kruegler+:2025}.
A third scenario, in which field-aligned electrical currents created by the accumulation of charge on the flanks of plasma injections give rise to injection signatures \citep{radioti+:2010}, analogous to the terrestrial case \citep{chen+wolf:1993}, has also been suggested, though the absence of inverted-V potential structures in the middle equatorial magnetosphere \citep{salveter+:2022} makes this scenario unlikely. 
In the scattering scenario, the electron population is expected to become more isotropic under the effects of pitch-angle scattering \citep{li+:2017} and thus enhancements in the field-aligned electron flux should be accompanied by an enhancement in the perpendicular electron flux.
Additionally, since the precipitating electron flux would be controlled by the loss-cone angle in the ionosphere, the corresponding injection signatures should emit less power in regions of the ionosphere where the magnetic field is stronger, since this reduces the loss-cone angle \citep{gerard+:2013}.
In the high-latitude Alfv\'{e}nic scenario, the electron acceleration is instead expected to be largely field-aligned \citep{saur+:2018} and thus not accompanied by perpendicular acceleration.
Alfv\'{e}nic acceleration is also expected to be more efficient at higher magnetic-field strength \citep{hess+:2013}, and thus, contrary to the scattering case, injection signatures should emit more power where the ionospheric magnetic field is stronger.
These characteristics provide us with metrics to determine to what extent these two acceleration mechanisms contribute to injection signatures.
\citet{gerard+:2013} observed that the north-south power ratio for injection signatures was neither directly nor inversely proportional to the surface field-strength ratio and were thus unable to differentiate between the two scenarios, though this was based on a single observation.

Discrete features between the main emission \citep[e.g.][]{grodent:2015} and the statistical Io footpath \citep[e.g.][]{bonfond+:2017,palmaerts+:2023} are observed at all System-III longitudes, at radial distances between 7 and 40 Jupiter radii (R$_{J}$), and have lifetimes greater than 45 minutes, compatible with \textit{Galileo} observations of plasma injections and first leading to their identification as injection signatures \citep{dumont+:2014}.
In addition, an increase in the brightness of injection signatures was observed when the main emission was expanded \citep{tao+:2018,nichols+:2017}, consistent with increased mass loading of the plasma torus, more iogenic plasma to be evacuated, and thus enhanced hot plasma injections to balance the magnetic-flux outflow \citep{bonfond+:2012}.
This link between internal processes and plasma injections is strengthened by the appearance of large injection signatures during periods of solar-wind quiescence \citep{kimura+:2016}.
Since 2016, \textit{Juno} traversals of plasma injections have permitted direct comparison with auroral morphology and the confirmation that discrete features in the outer emission are likely linked to plasma injections \citep{nichols+:2023}.
In addition to ``blob-like'' injection signatures, arc-like discrete features with isotropic electron distributions \citep{radioti+:2009} have also been identified that, while observed to enhance in brightness several days after plasma injection, were not directly identified as injection signatures \citep{gray+:2017}.

Injection signatures are often associated with dawn storms in the UV aurora \citep{bonfond+:2021}.
Dawn storms are very bright discrete features coincident with the dawn-side main emission which arise from dipolarisation in the dawn-side magnetosphere, provoked by reconnection in the magnetotail, which itself gives rise to plasma injections in the post-noon sector \citep{yao+:2020,bonfond+:2021}.
\citet{ebert+:2021} associated auroral dawn storms with plasma injections and field-aligned currents.
However, injections are only the final step of the dawn storm evolution track; an active dawn storm is more likely associated with dipolarisation rather than plasma injection \citep{bonfond+:2021}, which may not energise auroral electrons via the same mechanisms.
This transition from dawn storm has been previously reported from \textit{Hubble} Space Telescope data \citep{gray+:2016}.
However, in one case, an injection signature was observed to arise in the absence of any preceding dawn storm \citep{bonfond+:2017}.
This hints that not all injections are the consequence of the evolution of dawn storms, though the limited number of observed cases makes it difficult to draw concrete conclusions.
Indeed, quasi-constant small-scale-plasmoid or drizzle-like outflow have been suggested as mechanisms for outward mass transfer in the magnetosphere even in the absence of large-scale magnetotail reconfiguration \citep{bagenal:2007,bagenal+delamere:2011}.

The outstanding questions in the literature that this work aims to tackle are thus the following:
\begin{itemize}
    \item Can injection signatures be classed into two types (dawn-storm and non-dawn-storm) based on their behaviour and characteristics?
    \item Are injection signatures predominantly driven by scattering in the magnetodisc or high-latitude Alfv\'{e}nic acceleration?
    \item Is there a phenomenological difference between (blob-like) injection signatures and arcs in the outer emission?
\end{itemize}

\section{Observations}
\label{sec:observations}

Composite UV images from \textit{Juno}-UVS (68-210~nm; \citealt{gladstone+:2017_uvs}) are used in this work.
\textit{Juno}'s highly elliptical polar orbit allows it to view Jupiter's aurora in both hemispheres. 
Since \textit{Juno} is a spin-stabilised spacecraft, \textit{Juno}-UVS observes ``strips'' of Jupiter's aurora as the spacecraft rotates, which can be collaged to create maps of the aurora. 
For each pass of the \textit{Juno} spacecraft over Jupiter's poles (a perijove; e.g. PJ1-N for perijove 1, northern hemisphere), an exemplar map was produced by collating a 100-spin ($\sim$50-minute) stack of \textit{Juno}-UVS data that is centred as close as possible to the perijove time whilst covering at least 75\% of the auroral region \citep{bonfond+:2018}.
Radiation noise from the impact of high-energy electrons on the detector is also removed \citep{bonfond+:2021}.
A more detailed description of the production of this exemplar map is given in \citet{head+:2024}. 
This results in a representative view of the aurora in each hemisphere during each perijove, though care must be taken when interpreting these exemplar maps.
Different parts of the same feature may be sampled by spins that differ by 50 minutes, which may smear non-corotating auroral features.
When the fine detail of features is being investigated, or comparisons drawn between auroral morphology and instantaneous in-situ measurements, a 10-spin moving-window stack of \textit{Juno}-UVS maps is more appropriate; this is mentioned where relevant in the text.

In addition to image data from \textit{Juno}-UVS, data from the FluxGate Magnetometer (MAG-FGM; \citealt{connerney+:2017}) and the \textit{Juno} Energetic-particle Detector Instrument (JEDI; \citealt{mauk+:2017_jedi}) on board \textit{Juno} are used. 
Technical details of each instrument are contained within their associated reference.

\section{Methods}
\label{sec:corrections}


\begin{figure}[tbhp]
    \centering
    \includegraphics[width=0.5\linewidth]{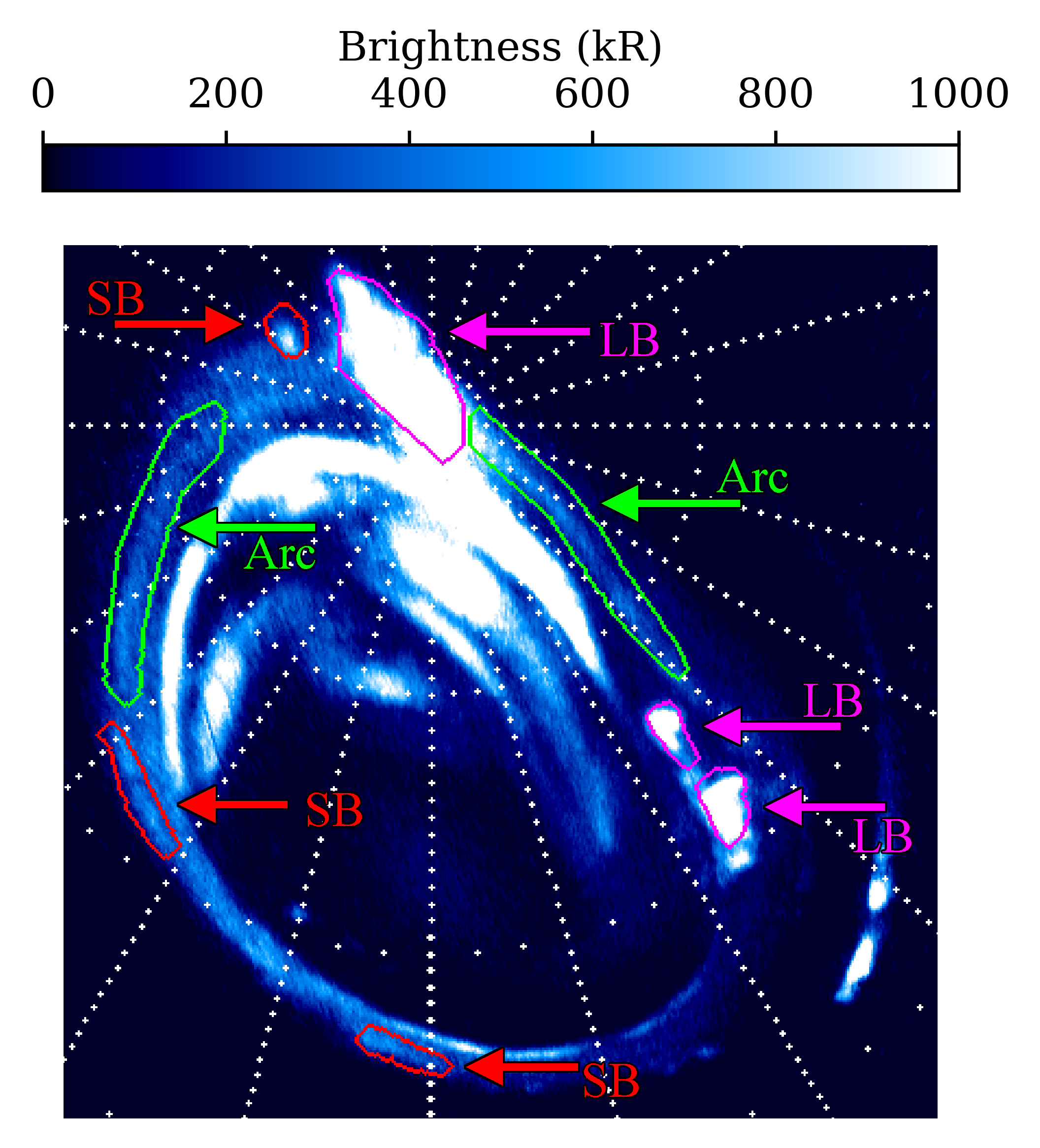}
    \caption{
        Exemplar map of UV brightness from PJ7-N, saturated to highlight injection signatures. Automatically detected features in the outer emission have been highlighted according to their feature type as defined in the text: red = small blob (SB), magenta = large blob (LB), green = arc. Gridlines are in jovicentric coordinates and are spaced by 15\textdegree\ in latitude and longitude.
    }
    \label{fig:feature_example}
\end{figure}

For this work, it is essential to have reliable estimates for the locations and shapes of injection signatures in UVS images of the aurora.
To this end, an injection-detection pipeline was developed that uses manual designations of injection-signature locations to train a pixelwise random-forest classifier \citep{ho:1998}.
This benefits from the accuracy of the manual designations whilst incorporating the objectivity of the automatic classifier.
A full description of this pipeline is given in appendix \ref{sec:rf_method}.
These detected discrete features were further sub-divided into three categories based on their shape when projected from the ionosphere into the magnetosphere equatorial plane:
\begin{itemize}
    \item Small blobs: longitudinal extent $<$ 30\textdegree, radial extent $<$ 5 R$_{J}$;
    \item Large blobs: longitudinal extent $<$ 30\textdegree, radial extent $>$ 5 R$_{J}$;
    \item Arcs: longitudinal extent $>$ 30\textdegree, radial extent $<$ 5 R$_{J}$.
\end{itemize}
The radial-extent cutoff of 5 R$_{J}$ was chosen to coincide with the ``compact-structure'' latitudinal-width cutoff (3\textdegree; 1\textdegree\ latitude $\sim$ 1.75 R$_{J}$ in the outer emission) used in \citet{dumont+:2014}.
The longitudinal-extent cutoff was selected qualitatively to differentiate between obvious arcs and more compact structures.
A sensitivity analysis to support the use of this cutoff, in the context of the analysis surrounding Figure \ref{fig:location_histogram}, is given in appendix \ref{sec:sensitivity_analysis}.
A case showing the three types of feature morphology is given in Figure \ref{fig:feature_example}.

Perijoves 1 to 40 are considered in this work.
The evolution of \textit{Juno}'s orbit is such that passes over the northern aurora occur at increasing speed and passes over the southern aurora at increasing altitude toward later perijoves.
This results in a diminishing UVS coverage of the aurora in the north and a diminishing resolution in the south, both of which hamper the reliable identification of injection signatures. 
As such, perijove 40 represents a reasonable cutoff for this work.
In total, this corresponds to 78 exemplar maps of the aurora (section \ref{sec:observations}), one per hemisphere per perijove excluding PJ2, when \textit{Juno} was placed into safe mode.
Exemplar maps of auroral brightness $B$ in the non-absorbed (145 - 165 nm) band in Rayleigh are converted to maps of estimated emitted power $P$ in watts by determining the surface area subtended by each pixel $A$ and applying
\begin{equation}
    P = 10^{10}\cdot\frac{1}{4} \cdot \frac{hc}{|\lambda|} \cdot A \cdot B \cdot 4.4,
\end{equation}
derived from the definition of the Rayleigh, where $h$ is the Planck constant, $c$ the speed of light, and 4.4 an empirical factor to convert the non-absorbed brightness into the full UV brightness.
In this work, the non-absorbed band is preferred over the more typical Lyman-$\alpha$ band (155 - 162 nm) because of its wider spectral range and hence greater signal-to-noise, especially to detect dimmer injection signatures.
This estimation of emitted power in this wavelength band is very approximate and assumes that all UV photons have the same wavelength of $|\lambda| = 160$ nm.
In this work, emitted power is only ever considered as the ratio of emitted power in the northern and southern hemispheres, and so the accuracy of the conversion is not essential as long as it is consistent between hemispheres.


In addition to maps of the auroral UV brightness, colour-ratio maps are also used in this work.
The (methane) colour ratio is used to probe the average energy of precipitating electrons, which we take as the ratio of the intensity in the 155-162 nm and 135-140 nm bands, since this definition has been shown to be less sensitive to known calibration issues in \textit{Juno}-UVS (\citealt{vinesse+:2025}, in revision) than the typical (155-162)/(125-130) nm definition \citep{gustin+:2013}.
In any case, \citet{vinesse+:2025} indicate that the two colour ratios deviate mainly for very bright features (such as the main emission and dawn storms) and so the effect of this new colour ratio on our conclusions is expected to be slight, especially since the absolute value of the colour ratio is not used to draw conclusions in this work.

The auroral brightness at the \textit{Juno} footprint is determined separately from the exemplar maps to ensure that the instantaneous brightness of the aurora below \textit{Juno} is captured, since the exemplar maps represent a single snapshot of the aurora during \textit{Juno}'s traversal of each hemisphere.
For the peak brightnesses (embodied by the 90th-percentile brightness seen during \textit{Juno} crossings of a discrete feature) presented in Figures \ref{fig:downward_flux_vs_brightness}, \ref{fig:alfvenic_flux_vs_brightness}, and \ref{fig:currents_vs_brightness}, an additional geometric correction of $r^{2}$, where $r$ is the radial distance from the centre of Jupiter in R$_{J}$, has been applied.
This is to compensate for the effect that \textit{Juno}'s altitude has on peak brightness; while the loss of resolution at higher altitudes should not affect the total power of an injection signature, the peak brightness value will decrease as $r^{-2}$.

The downward (planetward), upward, and perpendicular electron energy flux is determined from JEDI data accessed via the JMIDL tool provided by John Hopkins University. 
The downward and upward fluxes take electrons within 20\textdegree\ of the local magnetic-field vector, whereas the perpendicular flux takes electrons in the 45-135\textdegree\ range.
This is to ensure that the loss cone is never sampled by the ``perpendicular'' electron flux, since a loss-cone angle of 45\textdegree\ corresponds to an altitude of $\sim$0.26 R$_{J}$, below the minimum \textit{Juno}-crossing altitude for the injection signatures presented in this work.
An approximate loss-cone angle of 20\textdegree\ is reasonable for \textit{Juno}'s low-altitude passes over the aurora \citep{mauk+:2017}, though the exact angle varies with altitude and local surface magnetic field.
Since it would be prohibitively time-consuming to download electron-flux data within the exact loss-cone angle for each timestamp during each perijove, a geometric correction is applied by multiplying the 20\textdegree\ electron energy flux by the ratio between the calculated loss-cone angle at each point during \textit{Juno}'s traversal and 20\textdegree.
This is a simple correction that does not take into account e.g. the potential inhomogeneity of the flux within the loss cone, but it is necessary to get reasonable estimates of the loss-cone electron flux from JMIDL data.
In any case, perfect estimation of the loss-cone electron flux is difficult due to the lack of sampling by the most-field-aligned JEDI channels \citep{mauk+:2017}.

Alfv\'{e}nic Poynting flux is determined from \textit{Juno}-FGM data extrapolated via magnetic-flux conservation to the assumed auroral altitude of 400 km to ensure that the flux within the flux tube is correctly compared between perijoves regardless of \textit{Juno}'s observing altitude, after the method presented by \citet{gershman+:2019}.
A bandpass frequency filter with limits of (0.2, 5) Hz is applied to high-resolution (64 Hz; \citealt{connerney+:2017}) \textit{Juno}-FGM data; these filter frequencies are chosen to investigate Alfv\'{e}n waves that could potentially give rise to auroral electron precipitation \citep{lorch+:2022}.
This filtered magnetic-field data is compared to the JRM33 magnetic-field model to determine the field residual $\delta B$ and hence estimate the Alfv\'{e}nic Poynting flux in this frequency range 
\begin{equation}
    F_{\text{Alf}}=\frac{\left(\delta B\right)^{2}c}{\mu_{0}},
\end{equation}
where $\mu_{0}$ is the permeability of free space, as per \citet{gershman+:2019}.

Magnetic-field-line tracing is performed using the 18th-order JRM33 internal-field model \citep{connerney+:2022} combined with the Con2020 external-field model \citep{connerney+:2020}, via routines made available by the \texttt{JupiterMag} Python library \citep{jupitermag,wilson+:2023}.

Field-aligned currents were calculated from FGM data compared against the above magnetic-field model and extrapolated via conservation of current to the assumed auroral altitude of 400 km as per \citet{al_saati+:2022}; a full description is given in appendix A.2 of \citet{head+:2025}.

\section{Results}

\subsection{\textit{Juno}-UVS analysis}

\begin{figure}[tbhp]
    \centering
    \captionsetup[subfigure]{width=0.48\linewidth}
    \subfloat[Large blobs.]{
        \includegraphics[width=0.5\linewidth]{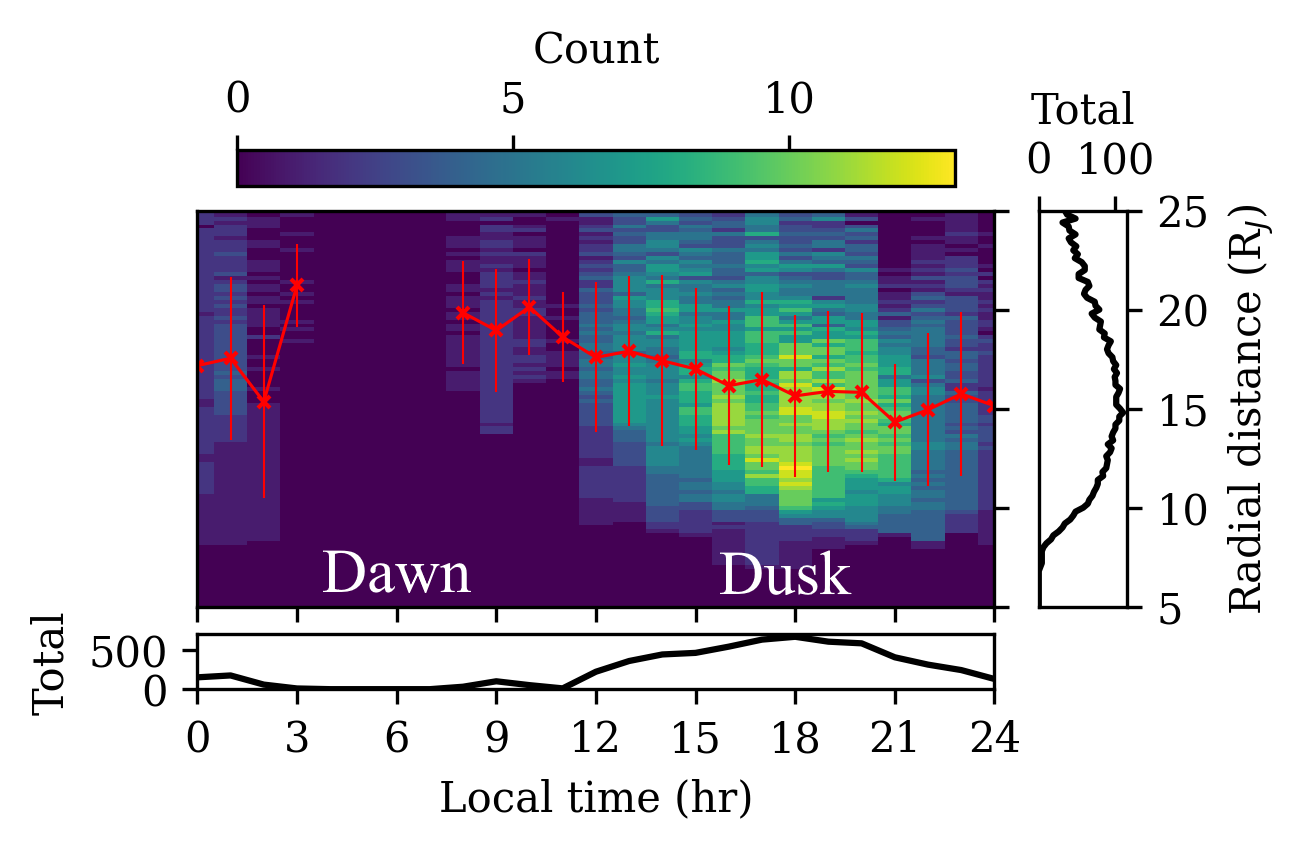}
    }\\
    \subfloat[Small blobs.]{
        \includegraphics[width=0.5\linewidth]{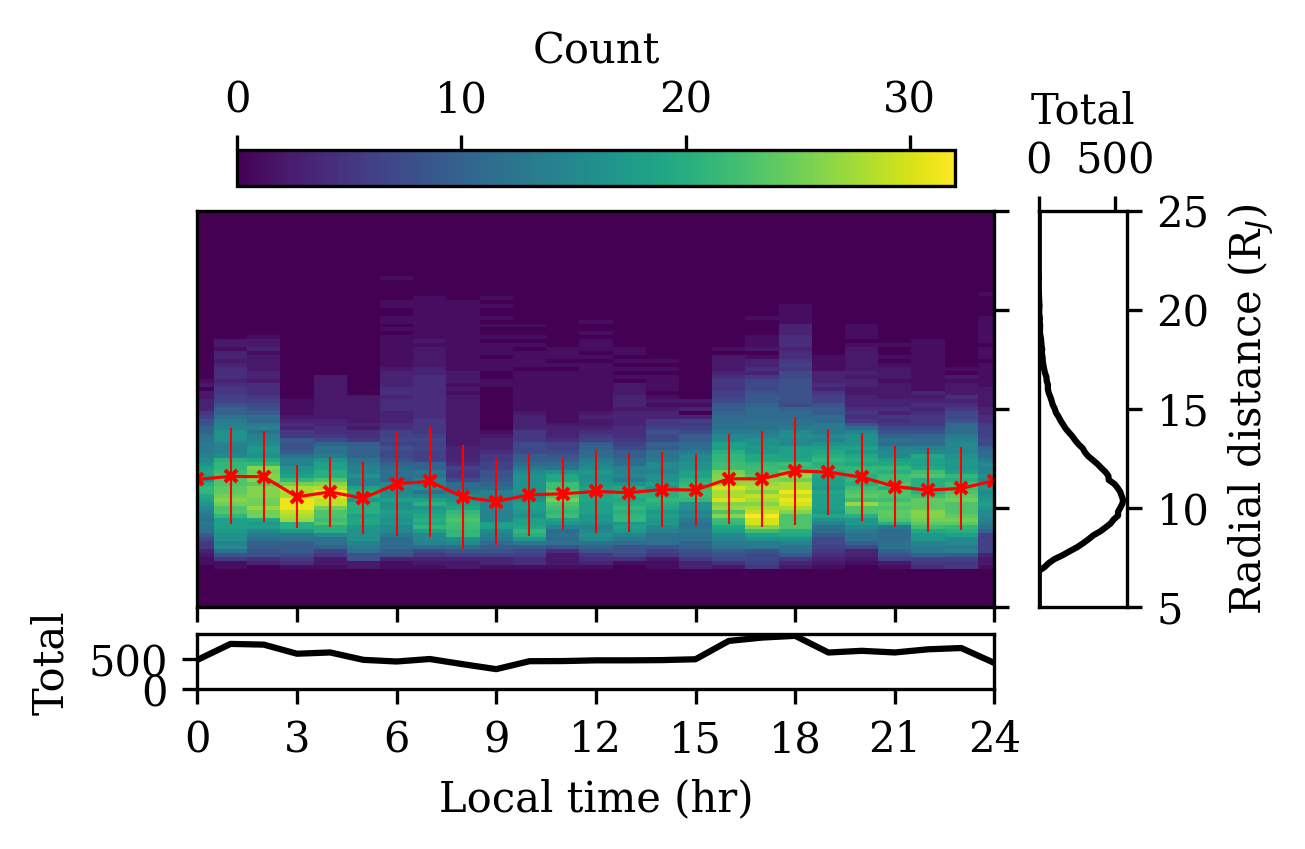}
    }\\
    \subfloat[Arcs.]{
        \includegraphics[width=0.5\linewidth]{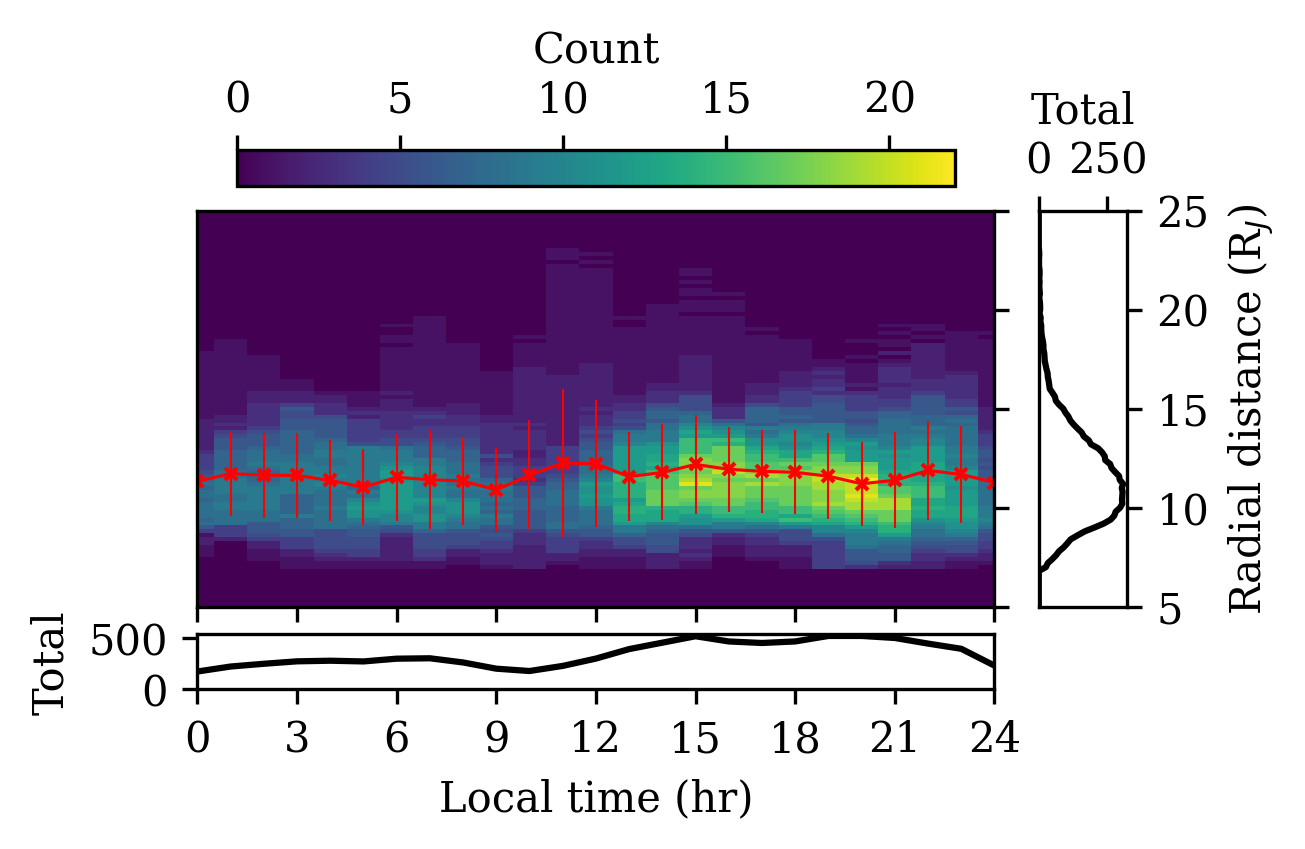}
    }
    \caption{
         Histogram in radial distance and local time of the projected position in the equatorial plasma sheet of features in the outer emission. The mean-average location for each local-time bin is given by the red line; error bars denote the standard deviation. Histograms flattened in local time and radial distance are given to the bottom and right of the main plot, respectively.  
    }
    \label{fig:location_histogram}
\end{figure}

Via magnetic mapping into the equatorial plasma sheet, the magnetospheric location of injection signatures can be investigated, as shown in Figure \ref{fig:location_histogram}.
Figure \ref{fig:location_histogram}a indicates that large-blob features predominantly map to the dusk-side magnetosphere between 10 and 20 R$_{J}$ from Jupiter.
This radial distance is compatible with the preferred distance for young plasma injections identified by \textit{Juno} ($\sim$17 R$_{J}$; \citealt{daly+:2024}).
The location of these large-blob features in the dusk-side magnetosphere is consistent with their interpretation as evolved dawn storms \citep{bonfond+:2021}; as dawn storms evolve, they move duskward and equatorward, transforming into injection signatures in the dusk-side outer emission.
This interpretation is strengthened by the fact that the average projected radial distance of the features decreases when moving from midday into dusk, as dawn storms leave the main emission and move equatorward in the midday sector \citep{bonfond+:2021}.
The absence of these features in the dawn-side outer emission therefore strengthens the link between large injection signatures and evolved dawn storms, which remain coincident with the main emission in the dawn sector and only transform into large injection signatures once they leave the main emission in the post-noon sector \citep{bonfond+:2021}.

As shown in Figure \ref{fig:location_histogram}b, small-blob features are observed much more uniformly at all magnetic local times (MLT), though there is still a slight preference for the dusk- and night-side aurora.
Additionally, their projected radial distance is closer to Jupiter than that of big-blob features, forming a dense population around 11 R$_{J}$ in agreement with earlier work \citep{dumont+:2014}.
It is curious that a significant portion of the small-blob population is located in the dawn side aurora, since dawn storms are expected to give rise to injections in the post-noon, dusk, and night sectors \citep{bonfond+:2021}.
This may be the result of a second class of injections that occur at all MLT.
Arc-like features in the outer emission show a very similar distribution as small-blob features (Figure \ref{fig:location_histogram}c), being concentrated around a radial distance of 11 R$_{J}$.
Their distribution in MLT, while more uniform than the large-blob injection signatures, is decidedly biased toward the dusk-side aurora when compared with the small-blob features.
The consequences of these results are discussed further in section \ref{sec:discussion}.


\begin{figure}[tbhp]
    \centering
    \captionsetup[subfigure]{width=0.48\linewidth}
    \subfloat[06:51:03]{
        \includegraphics[width=0.5\linewidth]{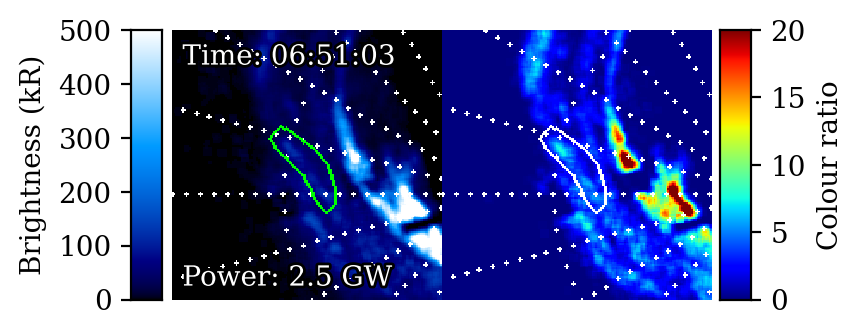}
    }\\
    \subfloat[07:17:19]{
        \includegraphics[width=0.5\linewidth]{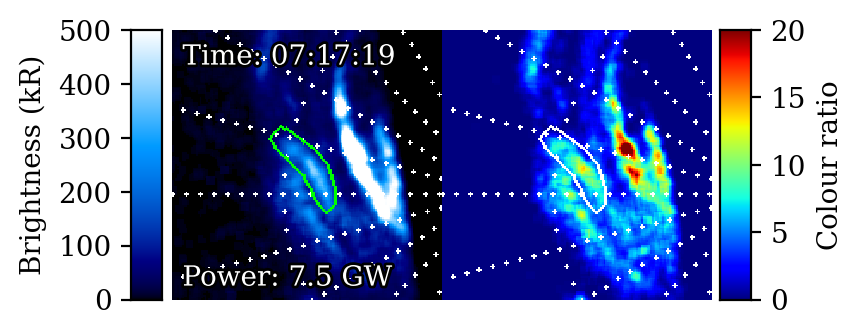}
    }
    \caption{
         An injection signature observed 2017-05-19 by \textit{Juno}-UVS during PJ6-S, highlighted in green. The brightness map is given on the left and the colour-ratio map on the right. The main emission is present to the right of the injection signature. 
    }
    \label{fig:spontaneous_example}
\end{figure}



In addition to the example given by \citet{bonfond+:2017}, this work identifies several further cases of injection signatures that appear in the absence of a preceding dawn storm; an example from PJ6-S is given in Figure \ref{fig:spontaneous_example}.
Here, at 06:51, the injection signature is barely present in the aurora, with a brightness comparable to the background brightness provided by the diffuse emission.
Later, at 07:17, the power of the injection signature has increased three-fold and now forms a clear, discrete structure distinct from the diffuse emission.
This example is present in the post-noon aurora ($\sim$13 MLT); two further examples in appendix Figures \ref{fig:spontaneous_example_PJ3} and \ref{fig:spontaneous_example_PJ7} are located at $\sim$9 MLT and $\sim$11 MLT respectively, indicating that non-dawn-storm injections may occur at a variety of local times.


\begin{figure}[tbhp]
    \centering
    \includegraphics[width=0.5\linewidth]{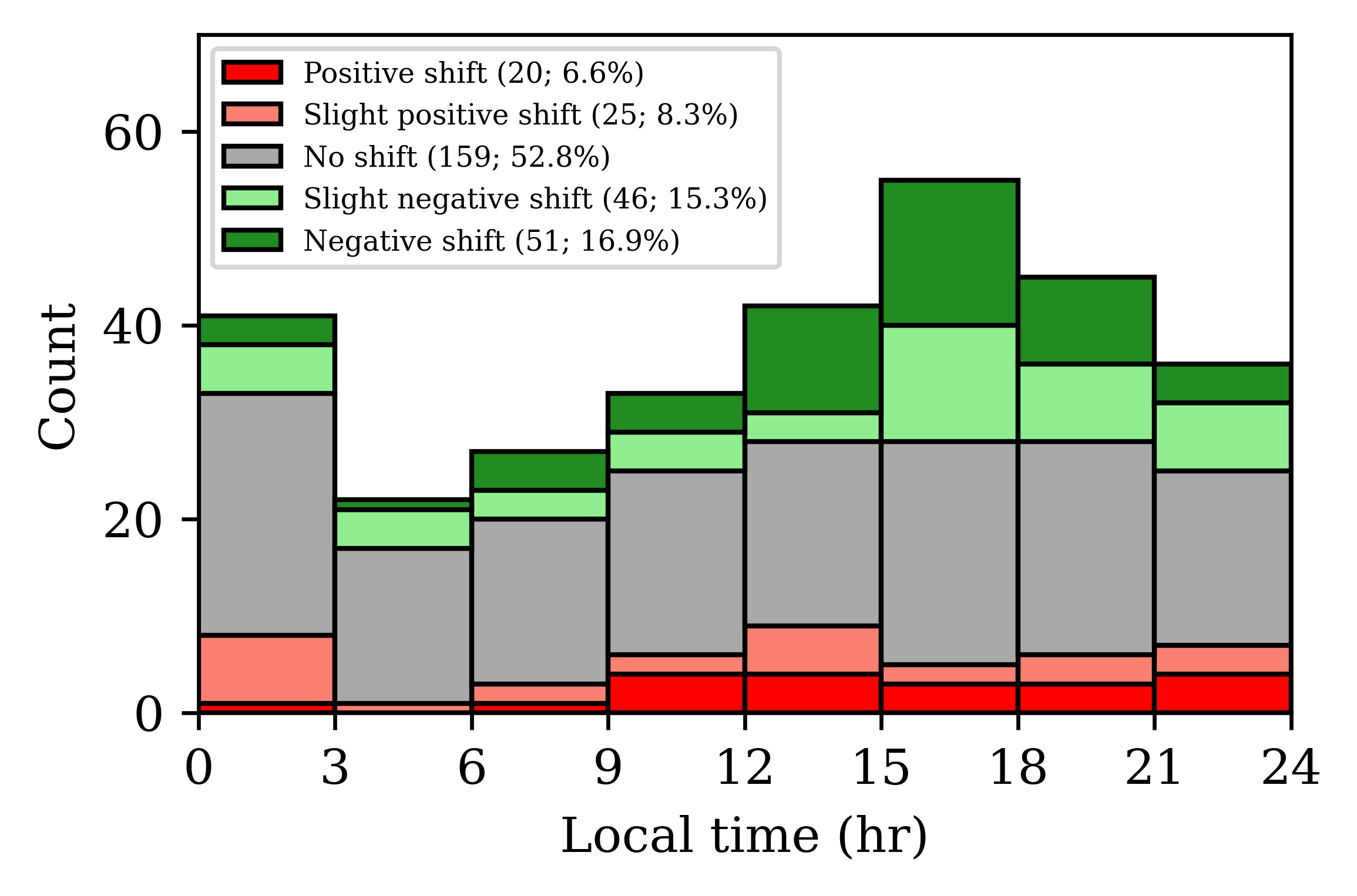}
    \caption{
Histogram in magnetospheric local time of blob-like injection features in the outer emission. Features are categorised by the magnetospheric longitude shift between the brightness peak and colour-ratio peak: no shift (<0.5\textdegree), slight shift (0.5\textdegree\ to 1\textdegree), and shift (>1\textdegree), from positive (red, bottom) through to negative (green, top) shifts.    
    }
    \label{fig:shift_histogram}
\end{figure}

Figure \ref{fig:shift_histogram} shows the shift between the brightness peak and colour-ratio peak for blob-like injection signatures.
A positive (negative) shift indicates that the brightness peak is at higher (lower) System-III longitude than the colour-ratio peak.
These shifts are calculated from the 10 highest-resolution (lowest-altitude) consecutive \textit{Juno}-UVS spins that captured at least 90\% of the injection signature; signatures with fewer than 10 consecutive spins are ignored.
This is to avoid spurious shift measurements in cases where the injection signature is not sufficiently sampled by \textit{Juno}-UVS, either spatially or temporally.
This method is preferred over the use of the exemplar maps to avoid introducing artefact brightness or colour-ratio peaks, since the exemplar maps are built from spins that may sometimes differ considerably in time.
Around half of detected injection signatures show no discernible shift between the brightness and colour-ratio peaks (absolute shift < 0.5\textdegree).
These features are relatively uniformly distributed in MLT, and a homogeneity test (a two-sided t-test) between the feature counts in the dawn sector (3 MLT to 12 MLT; this choice is motivated by the empty region in Figure \ref{fig:location_histogram}a where large-blob signatures are absent) and the dusk/night sector (12 MLT to 3 MLT) returns a p-value of 0.16. 
This is insufficient to indicate non-homogeneity at the 95\% confidence level.
Cases with noticeable shift have been separated into slight shifts (0.5\textdegree\ $\le$ shift < 1\textdegree) and considerable shifts (shift $\ge$ 1\textdegree) to increase the granularity of the analysis.
The negative-shift cases, which collectively account for 32\% of all detected features, appear to be concentrated in the dusk sector between 12 and 21 MLT; the same homogeneity test indicates a probability of 95\% that the negative-shift cases are not homogenous between the dawn and dusk/night sectors, which is a stronger indication of non-homogeneity than for the no-shift cases.
Positive-shift cases only account for 15\% of the detected injection signatures.
Of the 20 cases with strong positive shifts (>1\textdegree), 18 were found to be due to misdetection by the peak-finding algorithm, typically because the closest colour-ratio peak to the brightness peak was significantly smaller than a more distant peak (or vice versa), or because a single brightness peak presents as two slightly separated colour-ratio peaks (or vice versa).
This leaves only two signatures with strong (>1\textdegree) positive shifts that are not obviously the consequence of algorithmic failure.
In both of these cases, however, the widths of the brightness and colour-ratio peaks show significant overlap.
A similar inspection of the strong-negative-shift cases provides many more examples (36 out of 51 signatures, or 70\%) of convincing negative shift; some examples are given in Figures \ref{fig:shift_example_PJ19_N} and \ref{fig:shift_example_PJ19_S}.
This is as expected, since negative-shift cases correspond to injections where energy-dependent drift has forced the high-energy electrons (that give rise to the colour-ratio peak) ahead in System-III longitude of the lower-energy electrons (that, through a greater population, give rise to the brightness peak).
Positive-shift cases are therefore difficult to describe in this framework, since our current understanding supposes that energy-dependent drift always pushes the high-energy electrons ahead of the low-energy electrons \citep{mauk+:2002} due to the properties of the energy-dependent drift imposed by curvature and gradients in the magnetic field \citep{mauk+:1999,dumont+:2018}.
If we assume that all positive-shift cases are due to a deficiency in the algorithm, and that this deficiency works to introduce positive and negative artefacts symmetrically, we can subtract the positive-shift counts from the negative-shift counts to estimate the ``true'' distribution.
This returns a population of 75\% no-shift features, 9\% slight-negative-shift features, and 15\% negative-shift features. 
Thus, even after we compensate for the deficiencies of the peak-detection algorithm, negative-shift cases still make up a significant portion (24\%) of all injection signatures.
Note that exact proportions of each shift type are dependent on the implementation of the peak-detection algorithm; nevertheless, it can be said that negative-shift cases likely constitute a non-negligible portion of all injection signatures.


\begin{figure}[tbhp]
    \centering
    \includegraphics[width=0.5\linewidth]{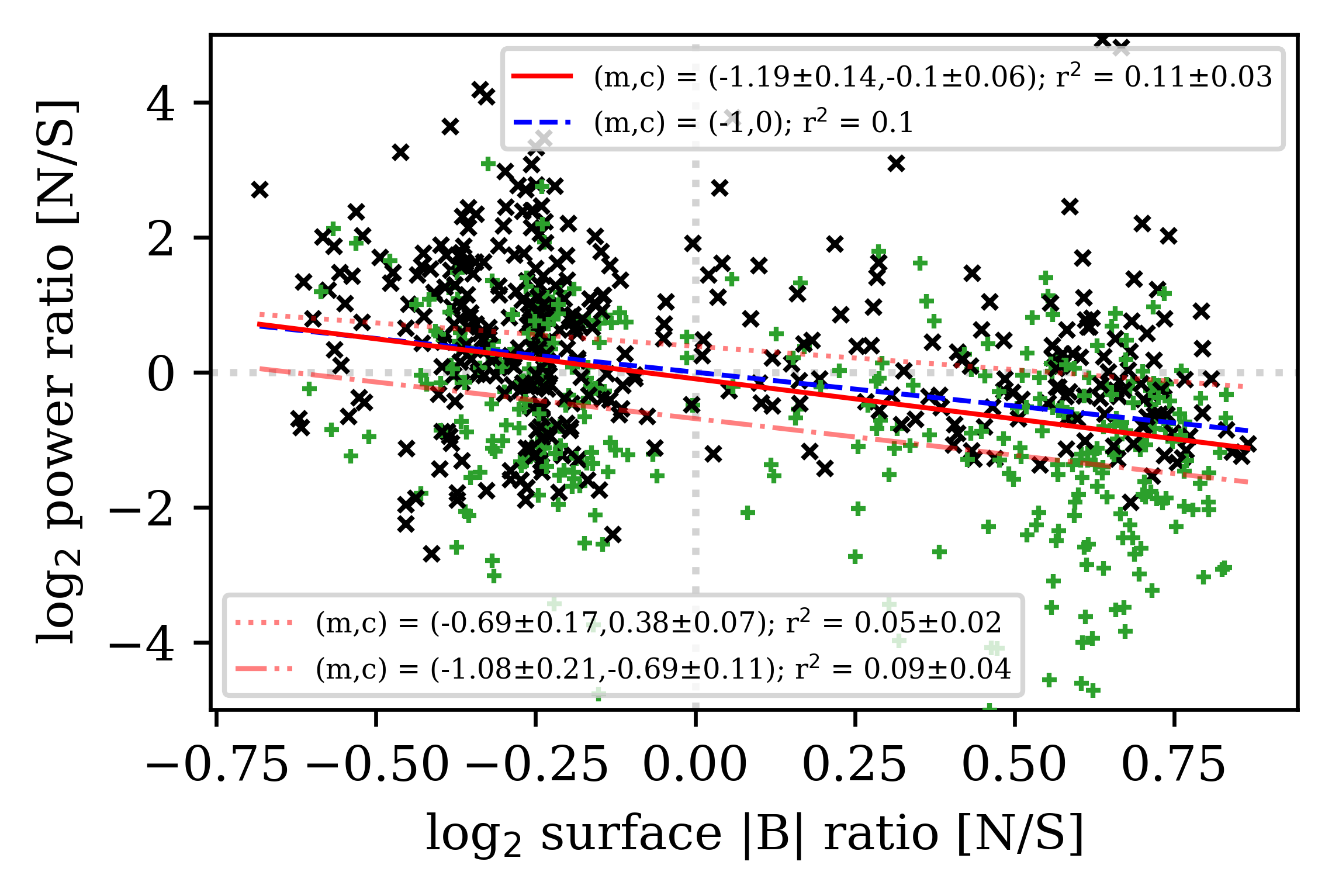}
    \caption{
        North-to-south UV auroral power ratio vs surface magnetic-field-magnitude ratio for injection signatures detected by \textit{Juno}-UVS during the first 40 perijoves. The N$\rightarrow$S projections are denoted by black $\times$, and the S$\rightarrow$N projections by green $+$. The best-fit linear relation for all points is given by a solid red line, and separate fitted relations for the N$\rightarrow$S and S$\rightarrow$N case by dotted and dash-dot lines respectively. The theoretical pitch-angle-scattering relation is given by a dashed blue line. 
    }
    \label{fig:field_power_ratio}
\end{figure}

One of the key observable differences between the magnetodisc-scattering and high-latitude-acceleration scenarios to precipitate electrons into injection signatures is the relation between injection-signature power and the local magnetic-field strength present for the injection signature within the ionosphere.
In the latter case, a greater precipitating electron flux is expected in regions of higher surface field strength, due to increased efficiency of the Alfv\'{e}nic acceleration process \citep{hess+:2013}, or more precisely
\begin{equation}
    \log_{2}\left(\frac{P_{N}}{P_{S}}\right) = -\log_{2}\left(\frac{B_{N}}{B_{S}}\right),
\end{equation}
following the derivation given in appendix \ref{sec:B_vs_power_derivation}, where $P$ is the emitted UV power, extrapolated from the power in the non-absorbed band as per section \ref{sec:corrections}, and $B$ the surface magnetic-field strength in the northern ($N$) and southern ($S$) hemispheres.  
Here, the ratio of auroral power between hemispheres is preferred to a direct comparison of auroral power with surface field strength to remove the influence of the intrinsic intensity of the plasma injection, the same for the injection signature in the northern and southern aurora, since this may otherwise mask the influence of surface field strength on injection-signature power. 
This analysis also supposes that conjugate injection signatures are sufficiently long-lived to remain visible in the southern hemisphere even after the approximately three-hour traversal of \textit{Juno} between hemispheres, which is supported by previous work \citep{dumont+:2018}.

This theoretical relation is plotted in Figure \ref{fig:field_power_ratio} alongside the observed north/south power ratios for the injection signatures analysed in this work.
Here, the pixel masks covered by the detected injection signatures (see appendix \ref{sec:rf_method}) in one hemisphere are projected along magnetic field lines to the other hemisphere, which creates a similar pixel mask in the other hemisphere.
The total projected power can be calculated from this projected mask and compared to the total power in the original mask, though this assumes that the injection is in full corotation and produces conjugate signatures in both hemispheres.
This means that most injections have two points in Figure \ref{fig:field_power_ratio} (N$\rightarrow$S, S$\rightarrow$N), but both points are included to account for cases where the injection signature is detected in only one hemisphere. 
It can be seen that the linear relation fitted to these points is very close to the theoretical scattering relation.
However, there exists significant scatter in the data points, as evidenced by the low R$^{2}$ value of 0.11.
Firstly, a t-test was performed with the null hypothesis that the power ratios of points with (negative log$_{2}$) field-strength ratios above 0 are higher than those below zero. 
This led to a rejection of the null hypothesis with a probability of 10$^{-13}$, indicating that the clustered population of points on the left of Figure \ref{fig:field_power_ratio} are strongly suggested to have greater power ratios than those points on the right.
This is not incompatible with the large scatter in Figure \ref{fig:field_power_ratio}; though the data in the left and right clusters show reasonable overlap in their (presumed Gaussian) power-ratio distributions, these distributions are sufficiently well-sampled by the data to say with considerable certainty that data on the right are taken from a population with a lower average power ratio than the data on the left. 
Secondly, injections are known to slightly sub-corotate \citep{dumont+:2018}, and so will have moved slightly between \textit{Juno}'s northern and southern pass.
This may explain why the the power contained within the other-hemisphere-projected injection-signature mask is consistently lower than the power within the original mask, regardless of in which hemisphere the original injection signature is detected; an injection signature may move enough between \textit{Juno}'s northern and southern pass that a part of it lies outside the projected mask, reducing the total power within the mask.
Taking a representative corotation of 85\% \citep{dumont+:2018}, an injection signature will have moved by $\sim$14\textdegree\ between \textit{Juno}'s northern and southern pass, which may be sufficient to explain the $\sim$50\% scatter ($\sim$1 on the y-axis of Figure \ref{fig:field_power_ratio}) for injection signatures typically of comparable longitudinal extent.
This possibility is investigated more rigorously in appendix \ref{sec:subcorotation} where it is shown that, for a typical injection signature, a slight sub-corotation of 80-90\% may largely explain the distribution of points in Figure \ref{fig:field_power_ratio}. 
These analyses increase confidence that the fitted relation in Figure \ref{fig:field_power_ratio} reflects physical reality, though the extent to which this scatter can be accounted for by the results of these analyses is difficult to estimate. 
It should be noted that, while these caveats exist for the interpretation of the fitted relation in Figure \ref{fig:field_power_ratio}, positive correlation between increased field-strength ratio and power ratio, and hence high-latitude Alfv\'{e}nic acceleration within injection signatures, is not supported by Figure \ref{fig:field_power_ratio}.

\subsection{\textit{Juno} multi-instrument analysis}

\begin{figure}[tbhp]
    \centering
    \includegraphics[width=0.5\linewidth]{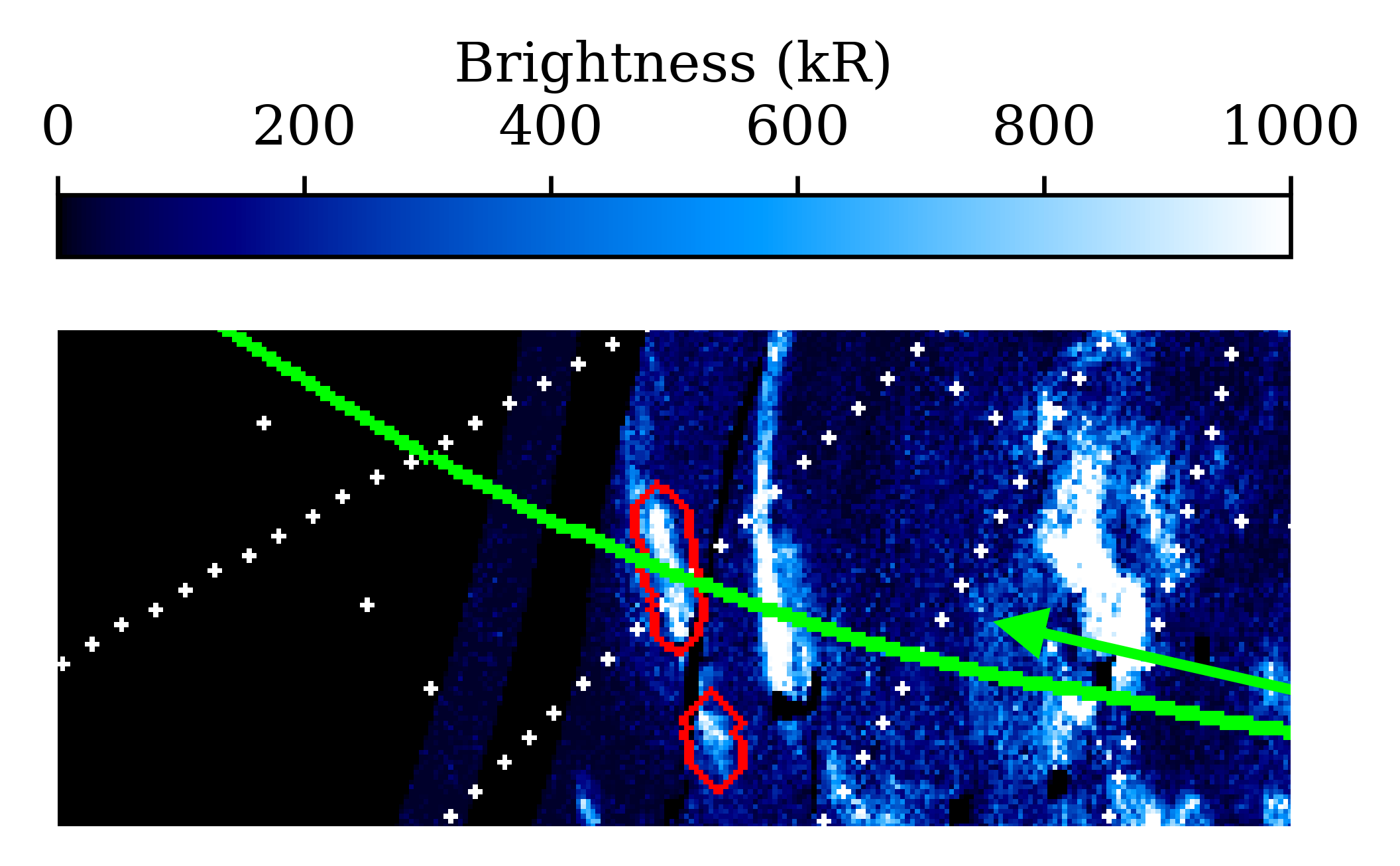}
    \caption{
        The \textit{Juno} footprint path (green) overlaid on the exemplar map of the aurora for PJ21-N. An automatically detected small-blob-type discrete feature crossed by \textit{Juno} is highlighted in red. The direction of travel of \textit{Juno} is denoted by the green arrow. 
    }
    \label{fig:crossing_PJ21_map}
\end{figure}

\begin{figure}[tbhp]
    \centering
    \includegraphics[width=0.5\linewidth]{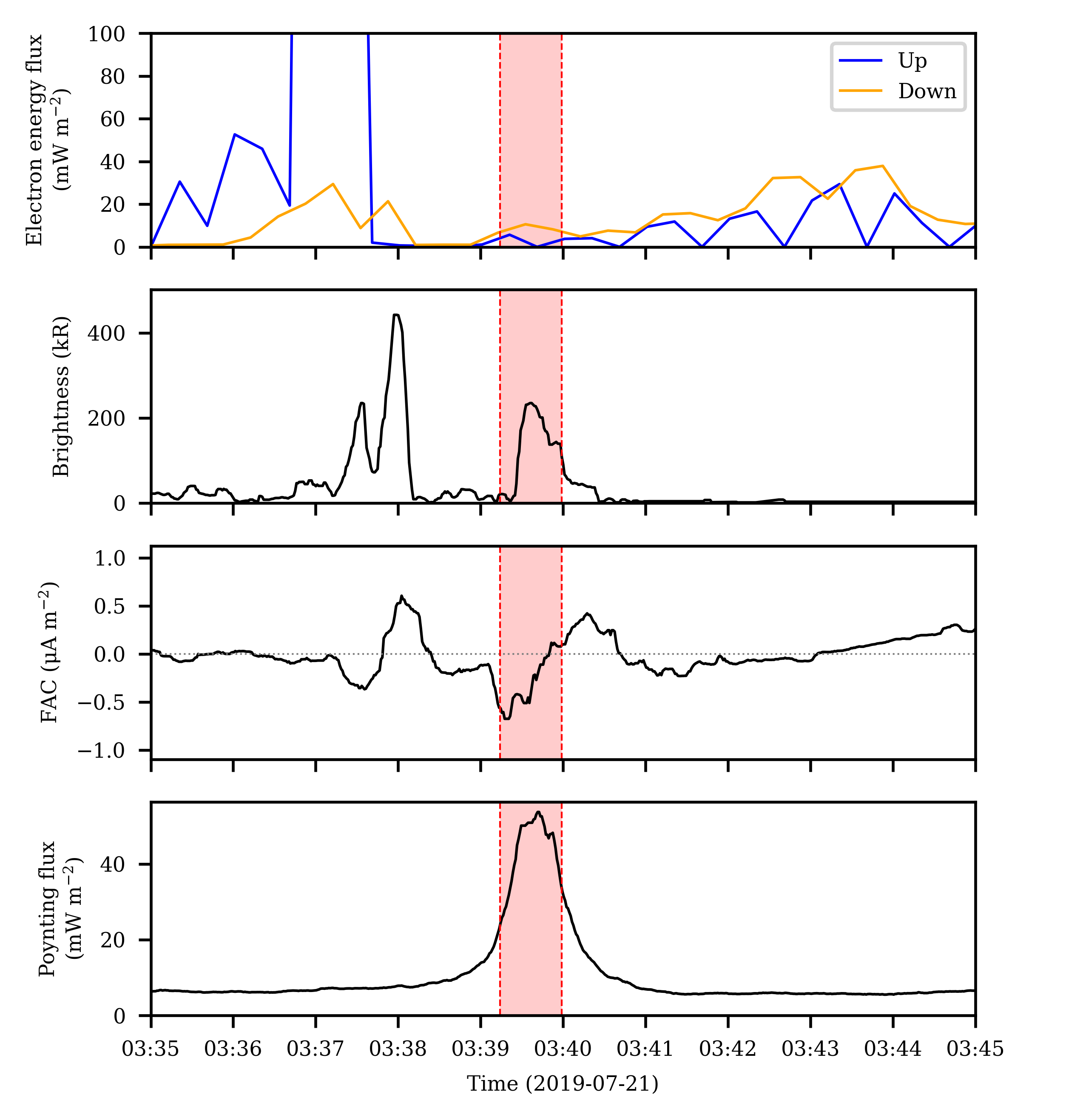}
    \caption{
        \textit{Juno} instrument data for PJ21-N. The crossing of the discrete feature in Figure \ref{fig:crossing_PJ21_map} is given in red. From top to bottom: JEDI field-aligned (0\textdegree-20\textdegree\ upward, 160\textdegree-180\textdegree\ downward) electron energy flux; UVS footprint brightness; calculated ionospheric field-aligned electrical current; calculated ionospheric Alfv\'{e}nic Poynting flux. 
    }
    \label{fig:crossing_PJ21_data}
\end{figure}

In addition to the analysis of auroral maps made by \textit{Juno}-UVS, the properties of injection signatures can be investigated using other instruments on board \textit{Juno}.
Figures \ref{fig:crossing_PJ21_map} and \ref{fig:crossing_PJ21_data} show the evolution of several calculated parameters as the \textit{Juno} footprint passed through a blob-like injection signature during PJ21-N.
The footprint UVS brightness was observed to peak within the range attributed to the feature crossing, which increases confidence that this feature is being properly detected.
The field-aligned downward electron energy flux shows a small peak ($\sim$15 mW m$^{-2}$) during the feature crossing, which is slightly below the energy flux expected to produce an auroral brightness of $\sim$200 kR (20 mW m$^{-2}$; \citealt{gerard+:2016}).
The dissipative Alfv\'{e}nic flux also shows a clear peak far above the background level.
A similar case where \textit{Juno} instead flew over an arc-like feature (Figures \ref{fig:crossing_PJ13_map} and \ref{fig:crossing_PJ13_data}) shows very similar behaviour.

\begin{figure}[tbhp]
    \centering
    \includegraphics[width=0.5\linewidth]{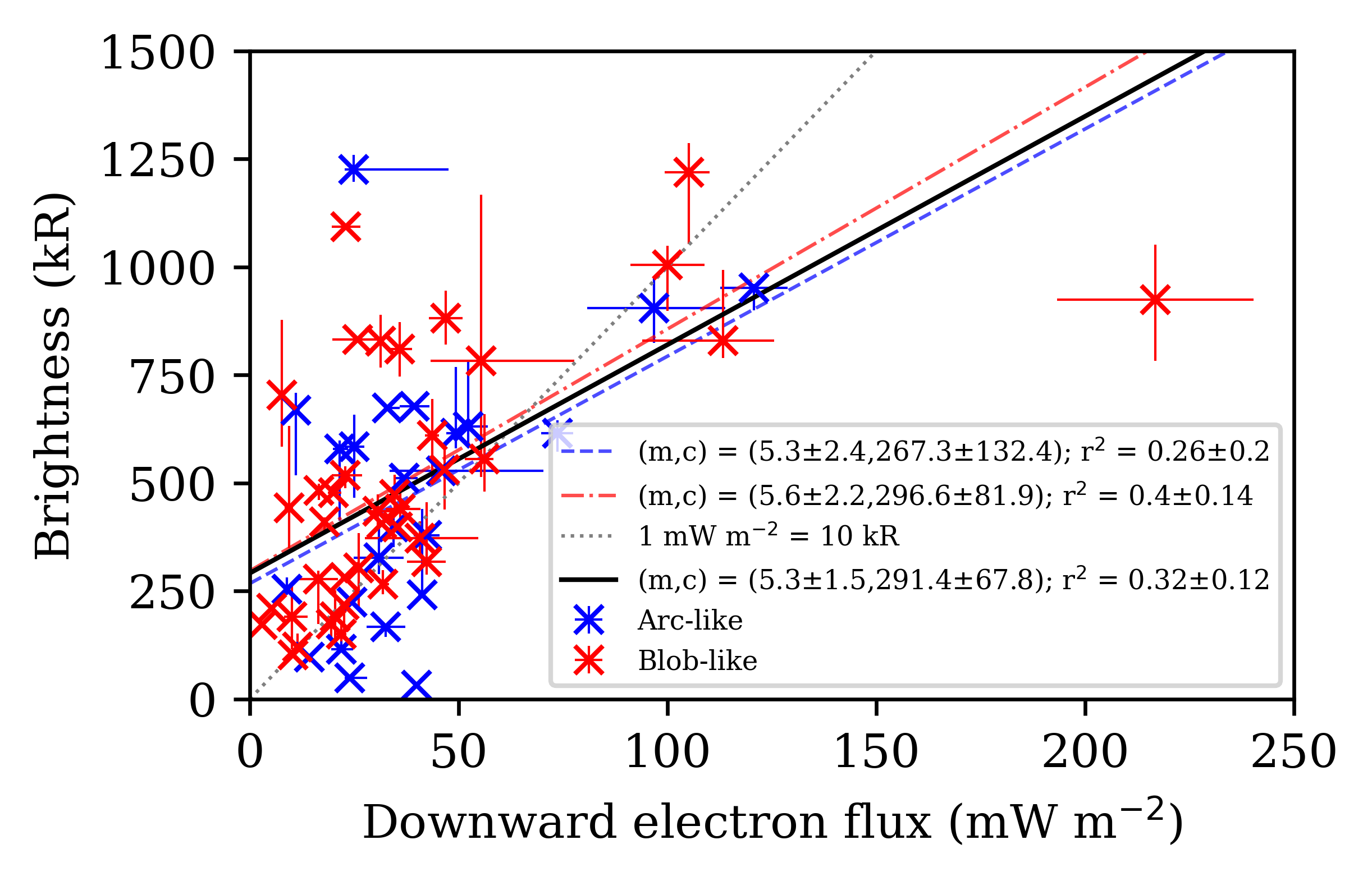}
    \caption{
        The 90th-percentile brightness vs 90th-percentile precipitating energy flux based on downward JEDI electron observed by \textit{Juno} during crossings of arc-like (blue) and blob-like (red) features in the outer emission. Error bars denote the 80th-100th percentile range. The best-fit linear relation is given by a solid black line, as well as separate best-fit relations for the arc-like (dashed) and blob-like (dash-dot) features. The theoretical relation after \citet{gerard+:2016} is given by the dotted grey line. 
    }
    \label{fig:downward_flux_vs_brightness}
\end{figure}

When considering \textit{Juno} crossings of both blob-like and arc-like discrete features in the outer emission during the first 40 perijoves, there exists a noticeable correlation between the downward electron-energy flux measured by JEDI and the instantaneous \textit{Juno}-footprint auroral brightness, as given by the solid black line in Figure \ref{fig:downward_flux_vs_brightness}.
However, while high electron-energy flux ($\gtrsim$ 80 mW m$^{-2}$) can be robustly associated with high auroral brightness ($\gtrsim$ 800 kR), the R$^{2}$ value of only 0.32 does not strongly support the fitted linear relation, and there exist several injection-signature crossings where high auroral brightness was coincident with only modest downward electron flux.
Several mitigating factors exist that may partially explain this discrepancy.
Firstly, the most field-aligned electrons are often poorly sampled by the JEDI instrument \citep{mauk+:2017}.
This can lead to an underestimation of the field-aligned electron energy flux, and one that is inconsistent between injection-signature crossings.
Secondly, while, for a typical average electron energy of 100 keV, the expected relation between electron energy flux and auroral brightness is 1 mW m$^{-2}$ $\sim$ 10 kR, this relation can vary slightly with electron energy \citep{gustin+:2016}, which may work to disrupt any linear relation that would otherwise be expected in Figure \ref{fig:downward_flux_vs_brightness}.
Finally, the loss-cone correction applied to the electron flux (described in section \ref{sec:corrections}), while necessary to extrapolate the electron flux within 20\textdegree\ to the loss-cone electron flux at \textit{Juno}, may also introduce some artificial error, especially when this relatively coarse correction is applied where the loss-cone angle differs significantly from 20\textdegree, though it should be noted that this correction improves the strength of the linear relation in Figure \ref{fig:downward_flux_vs_brightness} (R$^{2}$ = 0.11 $\rightarrow$ 0.32).  
With these caveats in mind, the principle that an increased downward electron energy flux leads to a higher auroral brightness is somewhat supported by Figure \ref{fig:downward_flux_vs_brightness}.
Figure \ref{fig:downward_flux_vs_brightness} also shows no meaningful separation between blob-like and arc-like discrete features in the outer emission.
Both morphologies occupy the same regions of the plot, and the fitted linear relations are identical within the calculated uncertainties.
This indicates that blob-like and arc-like outer-emission features have similar typical brightness/electron-flux ratios and hence a similar range of typical electron energies.

\begin{figure}[tbhp]
    \centering

    \includegraphics[width=0.5\linewidth]{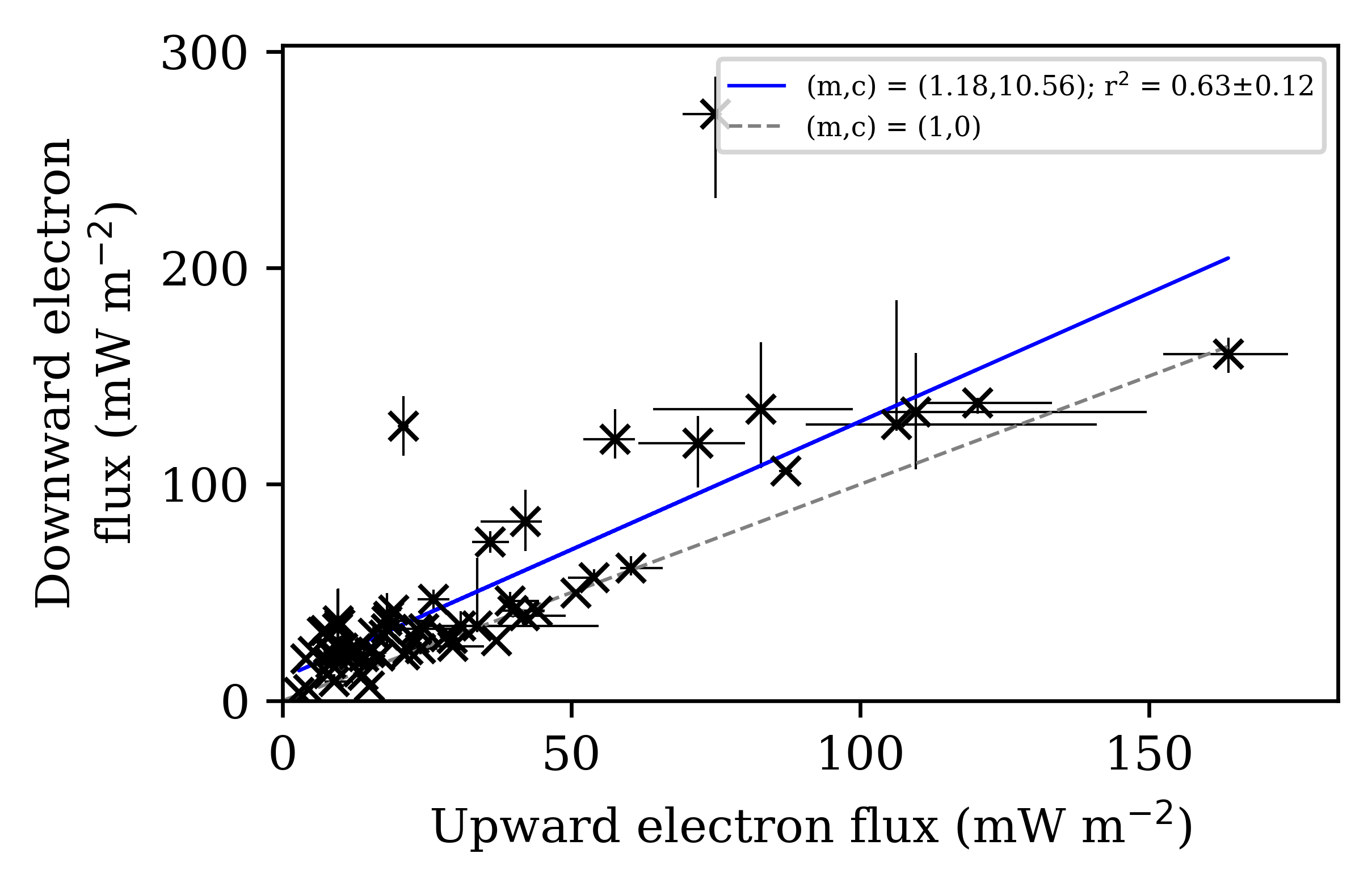}
    
    \caption{
        The 90th-percentile upward vs downward electron energy flux observed by JEDI during crossings of discrete features in the outer emission. Error bars denote the 80th-100th percentile range. The best-fit linear relation is given by a solid blue line, and the 1:1 relation by a dashed grey line.
    }
    \label{fig:downward_vs_upward}
\end{figure}

\begin{figure}[tbhp]
    \centering

    \includegraphics[width=0.5\linewidth]{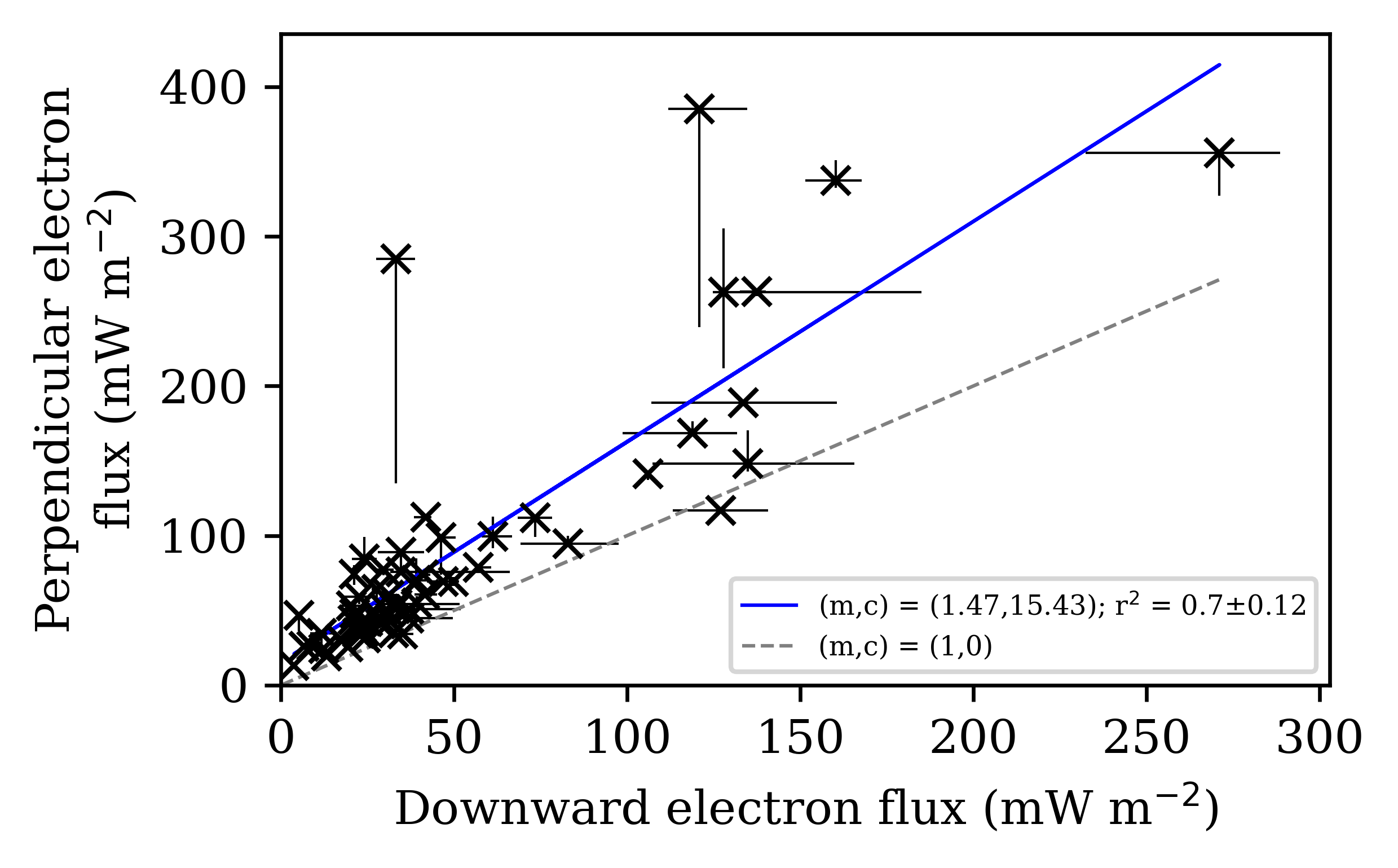}
    
    \caption{
        As Figure \ref{fig:downward_vs_upward} but for the downward vs perpendicular electron energy flux.
    }
    \label{fig:downward_vs_perp}
\end{figure}

As shown in Figures \ref{fig:downward_vs_upward} and \ref{fig:downward_vs_perp}, there exist also moderately strong linear relations between the upward and downward, and downward and perpendicular, electron energy fluxes.
Here, ``upward'' and ``downward'' refer to the electron population with 20\textdegree\ of the local magnetic-field vector, without the loss-cone correction as described in section \ref{sec:corrections}.
``Perpendicular'' electrons have pitch angles between 45\textdegree\ and 135\textdegree.
The lack of loss-cone-angle correction is done to compare the (fixed-angular-width) perpendicular electron flux with other parts of the pitch-angle profile, rather than to compare the auroral brightness with the precipitating electrons that give rise to it, as in Figure \ref{fig:downward_flux_vs_brightness}.
In any case, the effect of this correction on Figures \ref{fig:downward_vs_upward} and \ref{fig:downward_vs_perp} is minimal.
The reasonable correlation between all three portions of the directional electron flux indicates that the electron acceleration is likely to be, for the most part, isotropic and not simply bidirectional and field-aligned.
This is more compatible with magnetodisc scattering to provoke auroral electron precipitation \citep{li+:2017}, rather than high-latitude field-aligned acceleration, discussed further in section \ref{sec:discussion}.







\begin{figure}[tbhp]
    \centering
    \captionsetup[subfigure]{width=0.48\linewidth}
    \subfloat[Blob-like features.]{
        \includegraphics[width=0.5\linewidth]{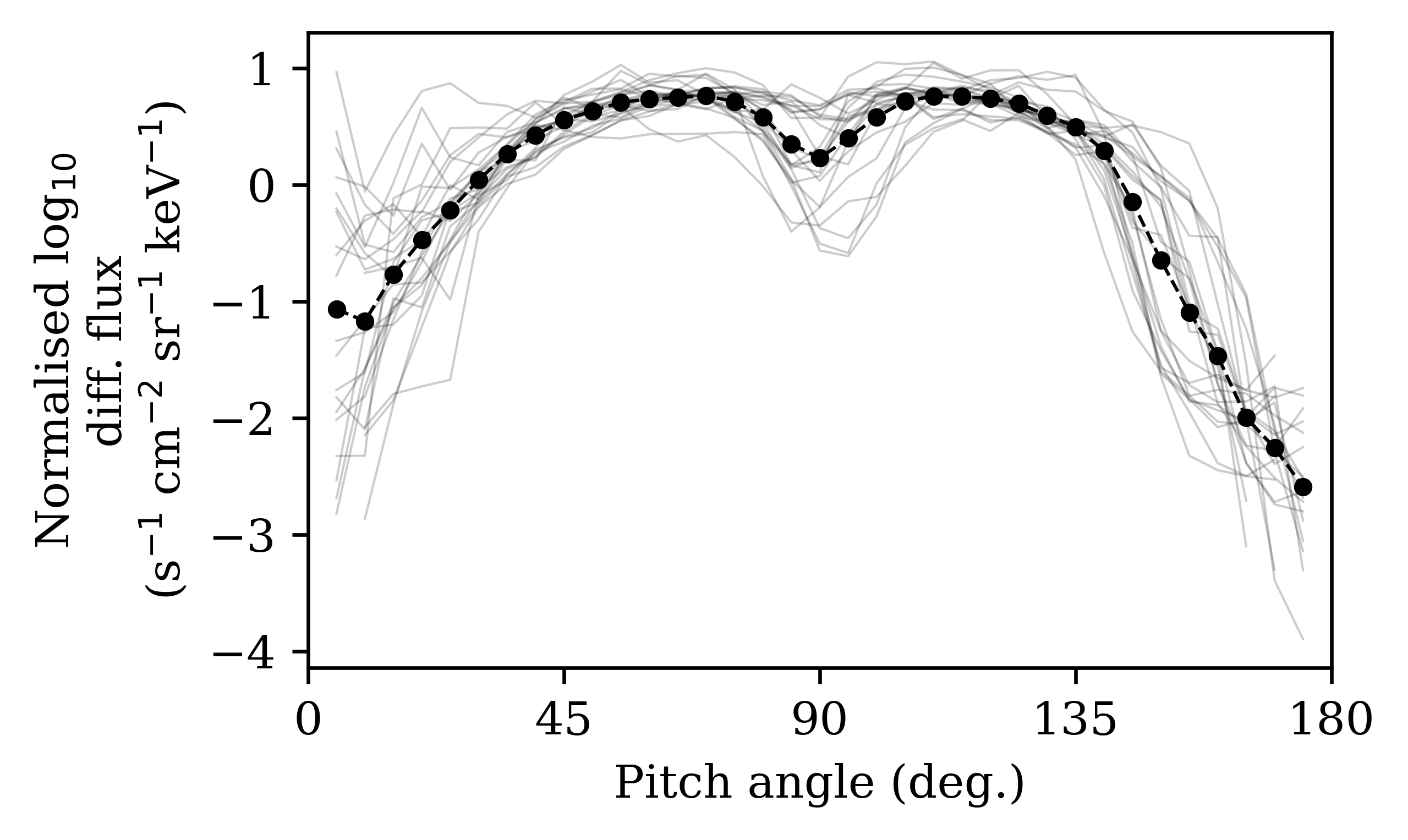}
    }\\
    \subfloat[Arc-like features.]{
        \includegraphics[width=0.5\linewidth]{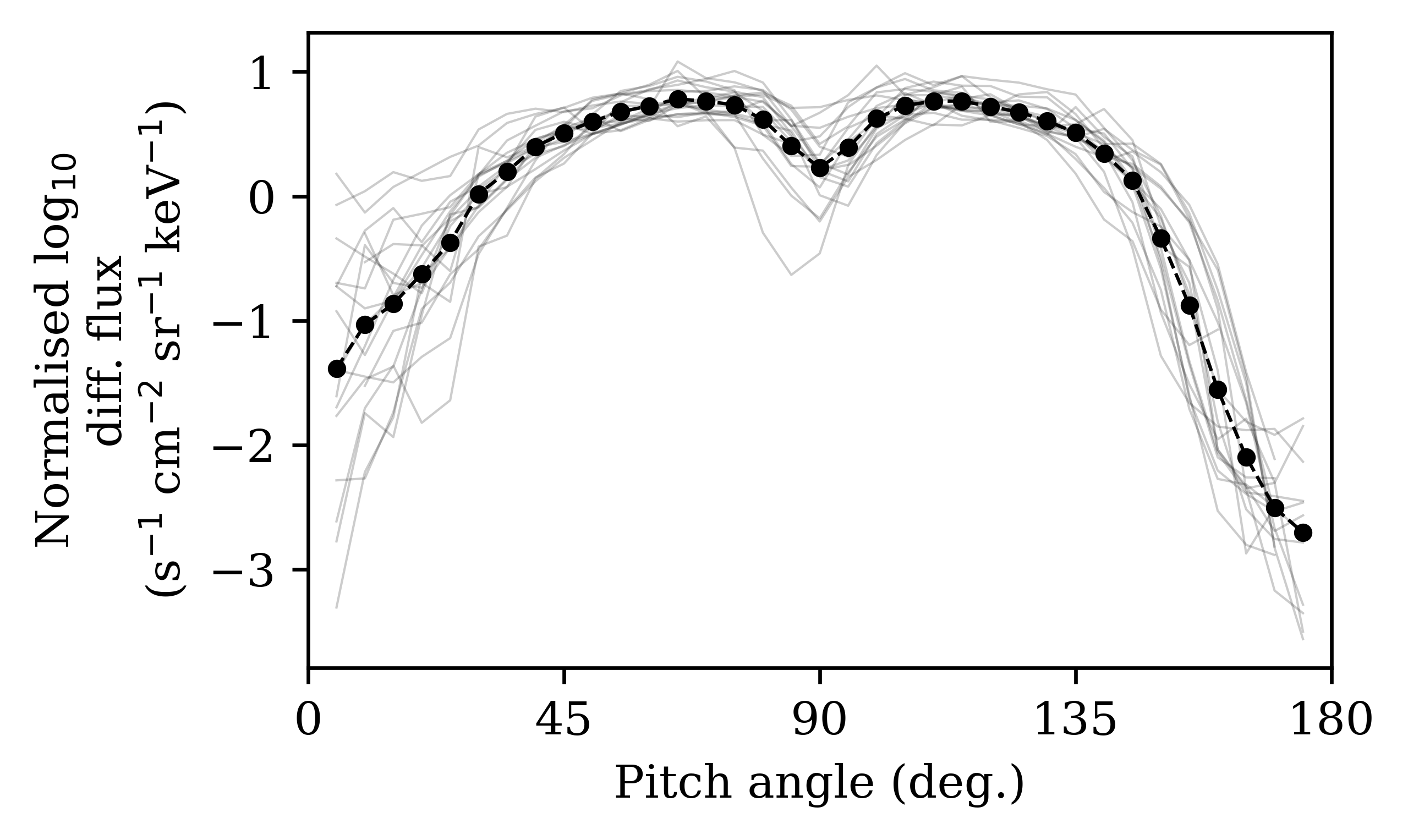}
    }
    \caption{
         The average electron pitch-angle distribution observed by JEDI during crossings of discrete features in the outer emission. Pitch-angle distributions from individual crossings are given in light grey. 0\textdegree\ (180\textdegree) denotes downward (upward) field-aligned electron flux.
    }
    \label{fig:pitch_angle_distributions}
\end{figure}

The full average electron pitch-angle profiles (in log$_{10}$ to highlight order-of-magnitude differences and normalised to prevent dominance of the greatest absolute electron flux in the average profile) for both blob-like and arc-like features (Figure \ref{fig:pitch_angle_distributions})  show butterfly distributions \citep{ma+:2017}, which indicate the same loss-cone depletion as pancake distributions \citep{salveter+:2022} but also include a ``notch'' at 90\textdegree. 
This notch is tentatively associated with a slight parallelisation of a pancake distribution via Landau damping of electrostatic waves at low altitude \citep{ma+:2017}. 
From Figure \ref{fig:pitch_angle_distributions}, it would appear that there is no significant difference in the average electron pitch-angle profiles of blob-like and arc-like features in the outer emission.
Additionally, the pancake-like butterfly distribution shown by both types of feature is indicative of partial refilling of the loss cones by isotropic scattering \citep{salveter+:2025}, which implies that the two corresponding source processes are at least comparable, if not the same process.

\begin{figure}[tbhp]
    \centering
    \includegraphics[width=0.5\linewidth]{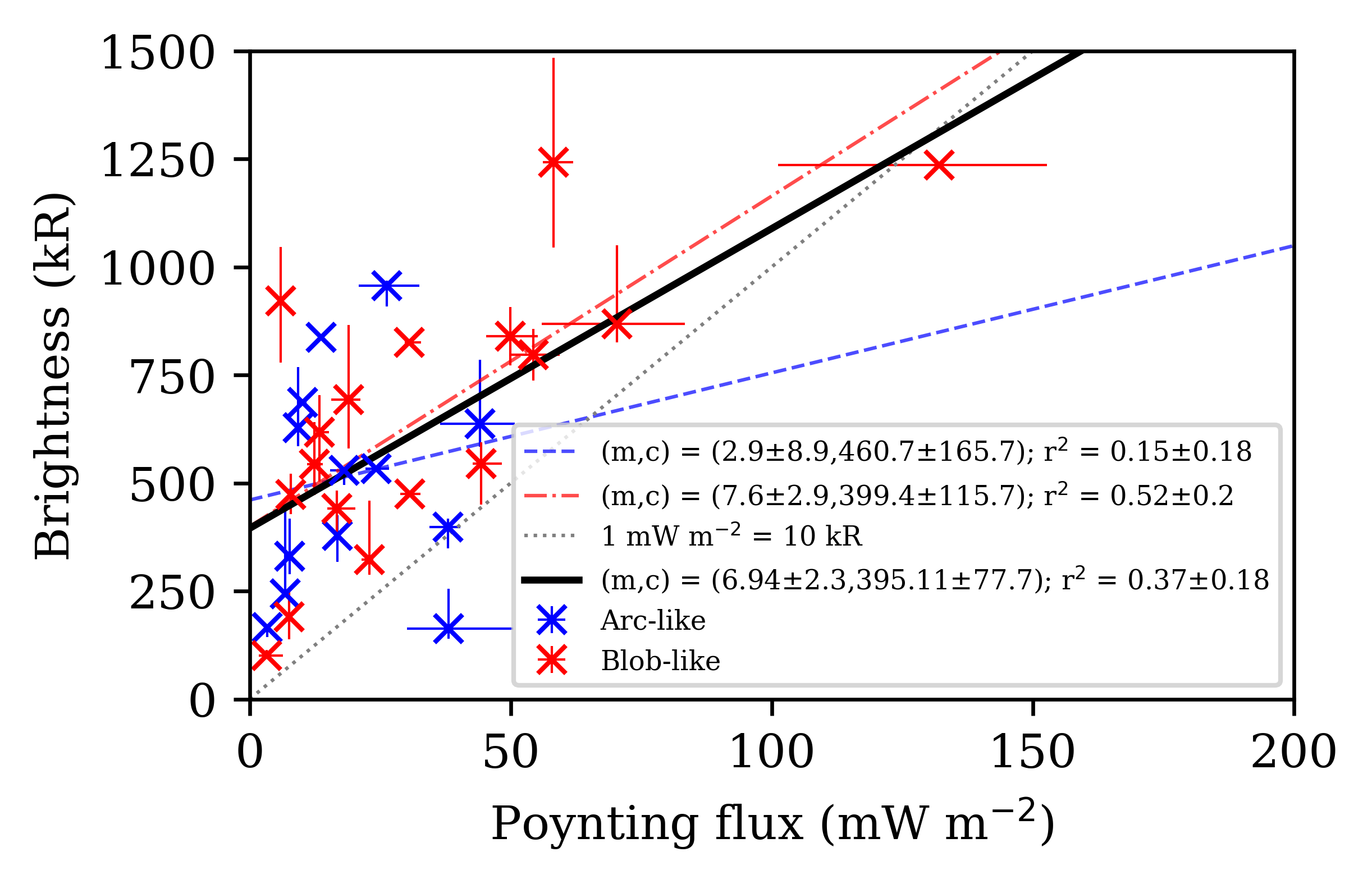}
    \caption{
        The 90th-percentile brightness vs 90th-percentile ionospheric Alfv\'{e}nic flux extrapolated from \textit{Juno}(-UVS, -FGM) observations below an altitude of 1 R$_{J}$ during crossings of arc-like (blue) and blob-like (red) features in the outer emission. Error bars and best-fit relations are denoted as per Figure \ref{fig:downward_flux_vs_brightness}.
    }
    \label{fig:alfvenic_flux_vs_brightness}
\end{figure}

Similarly, the Alfv\'{e}nic Poynting flux (in the potentially dissipative frequency range of 0.2 to 5 Hz; \citealt{lorch+:2022}) observed during \textit{Juno} crossings of injection signatures shows a reasonable linear correlation with auroral brightness, as in Figure \ref{fig:alfvenic_flux_vs_brightness}.
It should be noted that Figure \ref{fig:alfvenic_flux_vs_brightness} only includes cases where the \textit{Juno} crossing occurred below an altitude of 1 R$_{J}$, since no peaks in the Alfv\'{e}nic flux were observed above this altitude. 
Additionally, to ensure a sensible comparison between data points, cases where the \textit{Juno} crossing occurred in a region of magnetic field strength above 0.1 mT (where \textit{Juno}-FGM has a higher operational range than during the rest of a pass over the aurora; \citealt{connerney+:2017}) have been ignored.
Here, unlike in Figure \ref{fig:downward_flux_vs_brightness}, there appears to be a separation between arc-like and blob-like features, with the high-flux, high-brightness populated predominantly by blob-like features.
Given the lack of such a separation in Figure \ref{fig:downward_flux_vs_brightness}, it is anticipated that this is a result of the stricter filters for feature crossings in Figure \ref{fig:alfvenic_flux_vs_brightness} and low-number statistics rather than a physical effect.
In any case, there remains a reasonable (R$^{2}$ = 0.52) correlation between Alfv\'{e}nic Poynting flux below 1 R$_{J}$ and feature brightness for ``traditional'' blob-like injection signatures.
Since this Alfv\'{e}nic flux is still observable at low altitudes, and therefore not consumed by wave-particle interactions as expected for the main emission \citep{sulaiman+:2022}, we suggest that it does not contribute significantly to the electron precipitation associated with injection signatures and may instead be a by-product of the pitch-angle scattering process.

\begin{figure}[tbhp]
    \centering
    \includegraphics[width=0.5\linewidth]{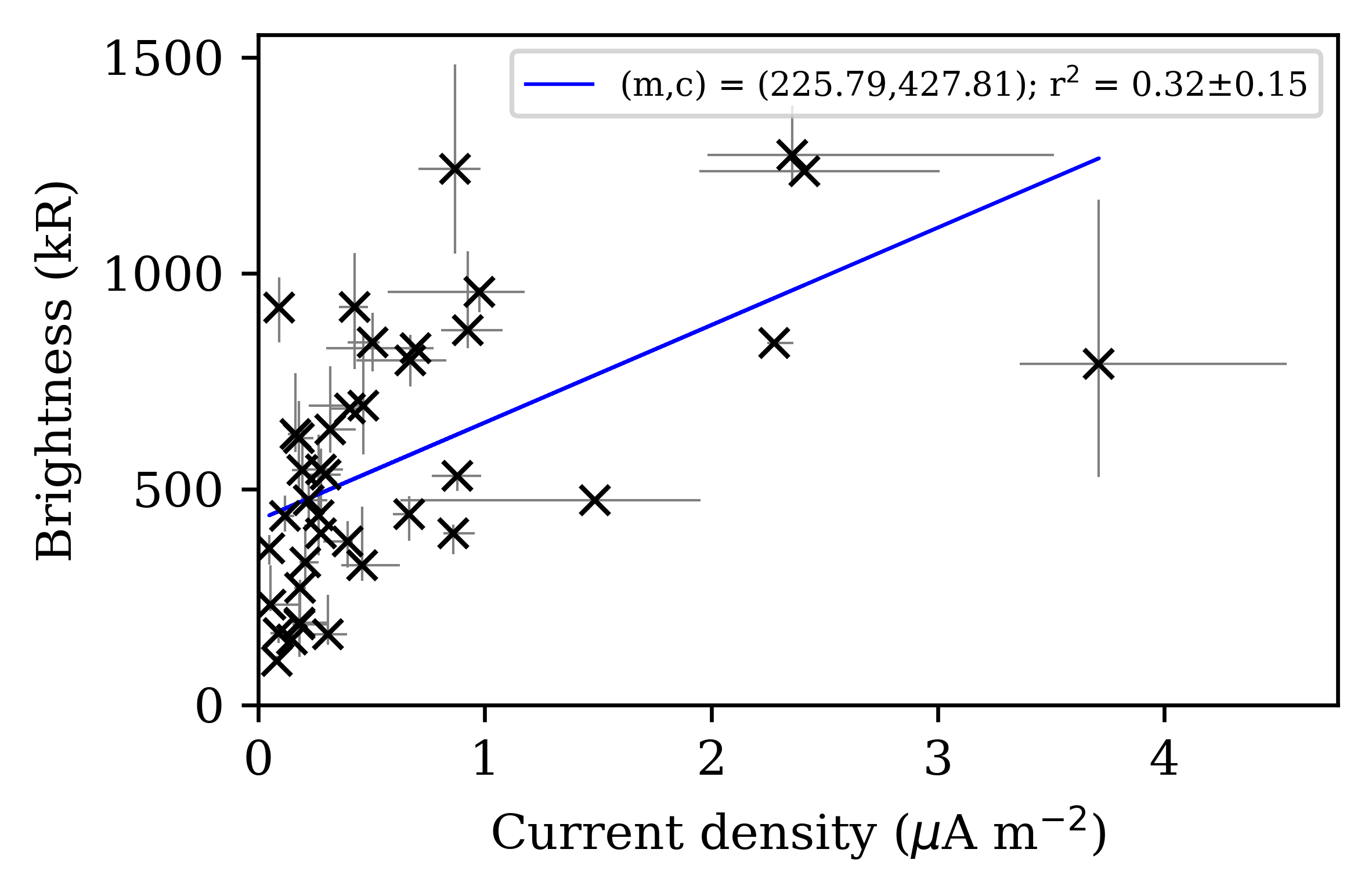}
    \caption{
        The 90th-percentile brightness vs 90th-percentile upward ionospheric current density extrapolated from \textit{Juno}(-UVS, -FGM) observations during crossings of discrete features below an altitude of 1 R$_{J}$ in the outer emission. Error bars denote the 80th-100th percentile range. The best-fit linear relation is given by a solid blue line.
    }
    \label{fig:currents_vs_brightness}
\end{figure}

Similar to the Alfv\'{e}nic flux, the magnitude of the upward electrical currents (that correspond to a majority of downward-travelling electrons) also show some measure (R$^{2}$ = 0.35) of correlation with observed auroral brightness for the discrete outer-emission features investigated in this work.
This is as expected for plasma injections; a parcel of plasma moving within a magnetic field will naturally lead to the production of electrical currents \citep{radioti+:2010}.
We may also expect that whichever property (or properties) of the plasma injection governs the brightness of the corresponding injection signature may also govern the intensity of the produced electrical currents.
However, similarly to the Alfv\'{e}nic flux discussed above, correlation between upward current density and auroral brightness does not necessarily imply a current-based origin for injection signatures, as discussed further in section \ref{sec:discussion}.

\section{Discussion}
\label{sec:discussion}

This work presents evidence that there may exist two classes of auroral injection signature, and hence two classes of injection.
The first, which we refer to as dawn-storm injections, is that of dawn storms which evolve through the large-blob phase and  become smaller injection signatures.
We would expect these injection signatures to be found in the post-noon, dusk, and night sectors \citep{bonfond+:2021} as the (sub-)corotating dawn storms move duskward as they evolve.
We would also expect these dusk-side injection signatures to show signs of ageing via energy-dependent electron drift \citep{bonfond+:2017}. 
The second is injection signatures that arise even in the absence of dawn storms.
The onset of a non-dawn-storm injection signature has been observed in only a single case prior to this work \citep{bonfond+:2017}, perhaps owing to their relatively short onset time compared to the typical lifetime of an injection signature. 
This work (Figures \ref{fig:spontaneous_example}, \ref{fig:spontaneous_example_PJ3}, \ref{fig:spontaneous_example_PJ7}) provides further examples of this phenomenon, indicating that the case presented by \citet{bonfond+:2017} was not a one-off event.
Indeed, the example given in Figure \ref{fig:spontaneous_example} is a small feature with no discernable shift between the brightness and colour-ratio peaks, indicative of a ``young'' injection.
We suggest that these smaller non-dawn-storm injection signatures arise at all MLT and therefore account for the considerable population of injection signatures in the dawn-side aurora; the dusk-side aurora thus consists of a mixture of the two types of injection signature.
This interpretation is consistent with the local-time distribution of injection signatures (Figure \ref{fig:location_histogram}).
Large-blob injection signatures are found exclusively at dusk because dawn storms evolve into injection signatures in this sector \citep{bonfond+:2021}, and small-blob injection signatures show a uniform distribution of non-dawn-storm injection signatures overlain by dusk-side dawn-storm injection signatures.
This is compatible with the distribution of brightness/colour-ratio shifts (Figure \ref{fig:shift_histogram}), which shows a uniform no-shift baseline for the ``young'' non-dawn-storm injections and a dusk-side concentration of negative-shift cases for the aged dawn-storm injection signatures.
The existence of non-dawn-storm injections at any local time may be consistent with a finger-like, interchange-dominated magnetodisc structure \citep{feng+:2023,yao+:2025} and/or the presence of reconnection events identified at all local times in the magnetodisc \citep{guo+yao:2024,zhao+:2024}.
Alternatively, the dawn-side population of no-shift injection signatures may be explained by the existence of significantly aged dawn-storm injections which have remained detectable and sub-corotated into the dawn-side aurora; in this scenario, the high-energy electron population may be completely depleted, which would bring the colour-ratio peak back in line with the brightness peak.
However, the threefold increase in auroral power over a timescale of hours for the dawn-side examples given in Figures \ref{fig:spontaneous_example}, \ref{fig:spontaneous_example_PJ3}, and \ref{fig:spontaneous_example_PJ7} is not easily compatible with significantly aged injection signatures arising from later-stage injections, since we would not expect these to increase in brightness at such a late stage in their evolution.
More work is required regarding the distribution, properties, and evolution of injections to be able to differentiate between these two scenarios.

The results of this work do not support high-latitude Alfv\'{e}nic acceleration as the dominant mechanism for the production of injection signatures. 
Firstly, the butterfly pitch-angle distributions (Figure \ref{fig:pitch_angle_distributions}) and the correlation between the downward, upward, and perpendicular flux in the low-altitude electron populations above injection signatures (Figures \ref{fig:downward_vs_upward} and \ref{fig:downward_vs_perp}) are consistent with an isotropic acceleration process, such as pitch-angle scattering in the magnetodisc, rather than a largely field-aligned process, as high-latitude Alfv\'{e}nic acceleration (and, indeed, acceleration via electrical currents; \citealt{ebert+:2021}) is expected to be.
This interpretation is strengthened by the lack of a positive correlation between hemispheric surface-field and auroral-power ratios for injection signatures, as would be expected for high-latitude Alfv\'{e}nic acceleration.
Indeed, Figure \ref{fig:field_power_ratio} shows some evidence for the inverse-log proportionality expected of isotropic scattering, though the dispersion in the data prevents us from drawing concrete conclusions from this analysis alone.
Finally, the existence of injection signatures with upstream drift of their colour-ratio peak (Figure \ref{fig:shift_histogram}) is far more consistent with the isotropic scattering of an energy-differentiated (aged) plasma injection than with high-latitude Alfv\'{e}nic acceleration \citep{dumont+:2018}.
\citet{dumont:2023} suggests that more no-shift (young) injection signatures are detectable at higher surface field strength in the north, whereas this distribution is more uniform in the south, which was suggested to be the consequence of Alfv\'{e}nic acceleration and its positive correlation between surface magnetic-field strength and auroral brightness.
We suggest instead that no-shift injection signatures are biased at high surface field strength (the maximum surface field strength is higher in the north) because a stronger surface field reduces the size of features in the ionosphere and thus makes it more difficult to identify non-zero brightness/colour-ratio peak shifts with the finite spatial resolution of \textit{Juno}-UVS.
It has also been suggested that modification of the local magnetic-field topology by a plasma injection may produce Alfv\'{e}n waves \citep{gray+:2017}, which may be the source of the Alfv\'{e}nic-flux peaks seen by \textit{Juno} during traversals of injection signatures.
However, since the evidence presented in this work indicates that pitch-angle scattering is the dominant acceleration process, it is unlikely that this Alfv\'{e}nic flux, which is expected to produce field-aligned acceleration, contributes significantly to electron acceleration above injection signatures.
Indeed, the fact that Alfv\'{e}nic flux in the potentially dissipative regime \citep{lorch+:2022} is observable at low altitude above injection signatures at all indicates that this flux does not significantly contribute to electron acceleration.
For example, the main emission, which is more frequently attributed to Alfv\'{e}nic acceleration in recent years \citep[e.g.][]{sulaiman+:2022,head+:2024,kruegler+:2025}, is not associated with significant Alfv\'{e}nic flux at low altitude \citep{sulaiman+:2022}; instead, dissipative Alfv\'{e}nic flux is observable at $\sim$10 R$_{J}$ \citep{lorch+:2022} which is then efficiently converted into electron acceleration at low altitude, resulting in bidirectional field-aligned electron distributions but no low-altitude Alfv\'{e}nic flux.
A similar scenario does not appear to be valid for injection signatures.
The moderate correlation between Alfv\'{e}nic flux and auroral brightness (Figure \ref{fig:alfvenic_flux_vs_brightness}) is perhaps indicative that the source process that isotropically scatters electrons in the equatorial plane also produces Alfv\'{e}nic perturbations, the strength of which scales with the strength of the scattering process.
Overall, the results of this work support pitch-angle scattering as the dominant mechanism for electron precipitation by plasma injections, consistent with previous work \citep[e.g.][]{li+:2017,dumont+:2018,devinat+:2025}.

Similarly, the correlation between upward electrical current density and auroral brightness shown in Figure \ref{fig:currents_vs_brightness} reminds us that correlation between two parameters does not imply causality.
Electrical currents, despite this correlation, are unlikely to be significant contributors to the electron precipitation associated with injection signatures; as discussed above, the ubiquity of electron pitch-angle distributions associated with isotropic scattering, as well the absence of inverted-V structures \citep{salveter+:2022}, above injection signatures is inconsistent with precipitation from currents induced by quasi-static electrical potentials.
We propose instead that plasma injections that give rise to injection signatures also produce field-aligned electrical currents due to charge accumulation on their flanks, as has previously been suggested \citep{radioti+:2010}, but that these currents do not result in significant auroral precipitation.

This work also presents evidence that blob-like and arc-like discrete features in the outer emission are both governed by the same fundamental processes.
Both feature types present comparable relations between downward electron flux/Alfv\'{e}nic flux and auroral brightness (Figures \ref{fig:downward_flux_vs_brightness} and \ref{fig:alfvenic_flux_vs_brightness}) and show near-identical average electron pitch-angle distributions (Figure \ref{fig:pitch_angle_distributions}).
Additionally, small blob-like and arc-like features are both projected to a preferred radial distance of $\sim$11 R$_{J}$ in the equatorial magnetosphere (Figure \ref{fig:location_histogram}).
This indicates that there is some measure of similarity between blob-like and arc-like features in the outer emission.
One key difference is that arc-like features seem to be less common in the dawn-side aurora.
We suggest that these results may be explained if arc-like features are interpreted as sequences of broadened or ``smeared-out'' blob-like features under the influence of energy-dependent electron drift, as is expected to occur \citep{dumont+:2018}.
Dawn storms are known to sometimes produce sequences of injection signatures in the post-noon sector \citep{bonfond+:2021}.
If these injections are sufficiently closely packed, energy-dependent electron drift may make it difficult to ascertain where one injection signature ends and the next begins.
Importantly, only dawn-storm injection signatures would be expected to show sufficient dispersion during the aged small-blob phase to broaden into an arc, since we anticipate that non-dawn-storm small-blob injection signatures would decay below the detection threshold before sufficient broadening had occurred; in any case, it is not presently clear why these non-dawn-storm injection signatures should occur in a sequence.
This interpretation explains why arc-like and small blob-like injection signatures preferentially occur at the same radial distance, but why arc-like signatures are less forthcoming in the dawn-side aurora, where dawn-storm injection signatures are absent.
The extended ``second auroral oval'' in the dawn-side aurora reported by \citet{gray+:2017} observed five days after a series of exceptionally bright injection signatures may be explained if the exceptional brightness of these sequential injection signatures allows them to remain detectable even once they have sub-corotated into the dawn aurora, at which point they would be expected to have undergone significant energy-dependent drift and hence have formed an extended arc.
The requirement for injection signatures to be initially exceptionally bright if they are to remain detectable once they reach the dawn-side aurora may explain the reduced (though still present) population of dawn-side arc-like features in Figure \ref{fig:location_histogram}c.

\section{Conclusions}
In response to the three questions tackled by this work, presented in the introduction, we present the following results:
\begin{enumerate}
    \item There is some evidence to suggest that injection signatures (and consequently injections) may be classified into two types: dawn-storm and non-dawn-storm. Dawn-storm injections occur primarily in the dusk and night sector and present as large, bright injection signatures that decay into smaller injection signatures that show ageing via energy-dependent electron drift. Non-dawn-storm injections can occur at any local time and present as small, unaged injection signatures.
    \item Injection signatures appear to be dominated by pitch-angle scattering of electrons in the magnetospheric equatorial plane rather than by field-aligned currents or Alfv\'{e}nic acceleration at high latitude. The low-altitude auroral electron energy flux increases isotropically with injection-signature brightness, and higher surface magnetic-field strength typically leads to a dimmer injection signature, consistent with pitch-angle scattering.
    \item Arc-like discrete features in the outer emission may consist of sequences of dawn-storm injection signatures that have undergone broadening via energy-dependent electron drift. \textit{Juno} crossings of both types of feature revealed no significant difference in their typical electron populations or Alfv\'{e}nic flux. They map to the same radial distance as small blob-like features but are more preferentially located in the dusk-side aurora, consistent with injection signatures resulting from dawn storms.
\end{enumerate}
Further work should aim to identify further examples of non-dawn-storm injection signatures and test whether they are truly uniformly distributed in local time, an analysis that is not possible at present due to the low number of conclusively identified non-dawn-storm injection signatures.
Confirmed \textit{Juno} crossings of non-dawn-storm injection signatures would also allow us to investigate whether the two classes of injection signature differ in their particle/wave/current properties, which would likely require a combination of in-situ measurement and remote observation.
A more extensive comparison of blob-like and arc-like discrete features, perhaps looking at the average energy of the precipitating electrons or an extended sequence of observations, may provide further evidence that arc-like features in the outer emission consist of sequences of broadened injection signatures.

\section*{Data availability statement}
\textit{Juno} data can be obtained from the NASA Planetary Data System (\url{https://pds-atmospheres.nmsu.edu/data_and_services/atmospheres_data/JUNO/juno.html}).

\section*{Acknowledgements}
   We are grateful to NASA and contributing institutions which have made the \textit{Juno} mission possible. This work was funded by NASA's New Frontiers Program for \textit{Juno} via contract with the Southwest Research Institute. This publication benefits from the support of the French Community of Belgium in the context of the FRIA Doctoral Grant awarded to L. A. Head. B. Bonfond is a Research Associate of the Fonds de la Recherche Scientifique - FNRS.  Data analysis was performed with the AMDA science analysis system provided by the Centre de Donn\'{e}es de la Physique des Plasmas (CDPP) supported by CNRS, CNES, Observatoire de Paris and Universit\'{e} Paul Sabatier, Toulouse. A. Moirano is supported by the Fonds de la Recherche Scientifique - FNRS under Grant(s) No T003524F. V. Hue acknowledges support from the French government under the France 2030 investment plan, as part of the Initiative d’Excellence d’Aix-Marseille Université – A*MIDEX AMX-22-CPJ-04, and CNES.

\bibliography{references}

\FloatBarrier

\begin{appendix}


\renewcommand{\thefigure}{A.\arabic{figure}}
\renewcommand{\thetable}{A.\arabic{table}}
\setcounter{figure}{0}
\setcounter{table}{0}
\setcounter{equation}{0}
\section{Description of feature-detection algorithm}
\label{sec:rf_method}

\begin{figure}[tbhp]
    \centering
    \includegraphics[width=0.5\linewidth]{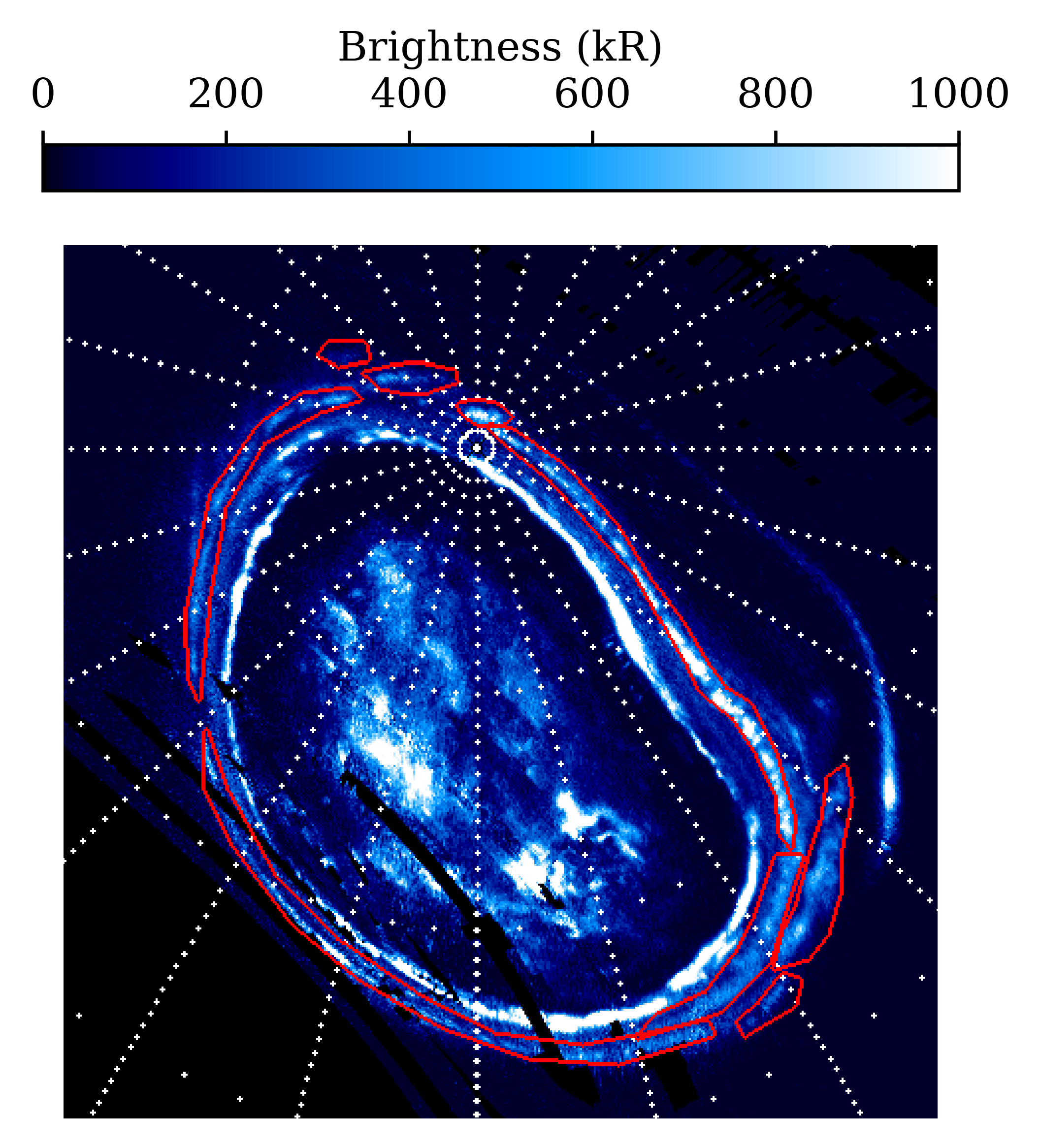}
    \caption{
        Exemplar map of UV brightness for PJ11-N. Manually designated features in the outer emission have been highlighted in red.   
    }
    \label{fig:rf_method_example_manual}
\end{figure}

\begin{figure}[tbhp]
    \centering
    \includegraphics[width=0.5\linewidth]{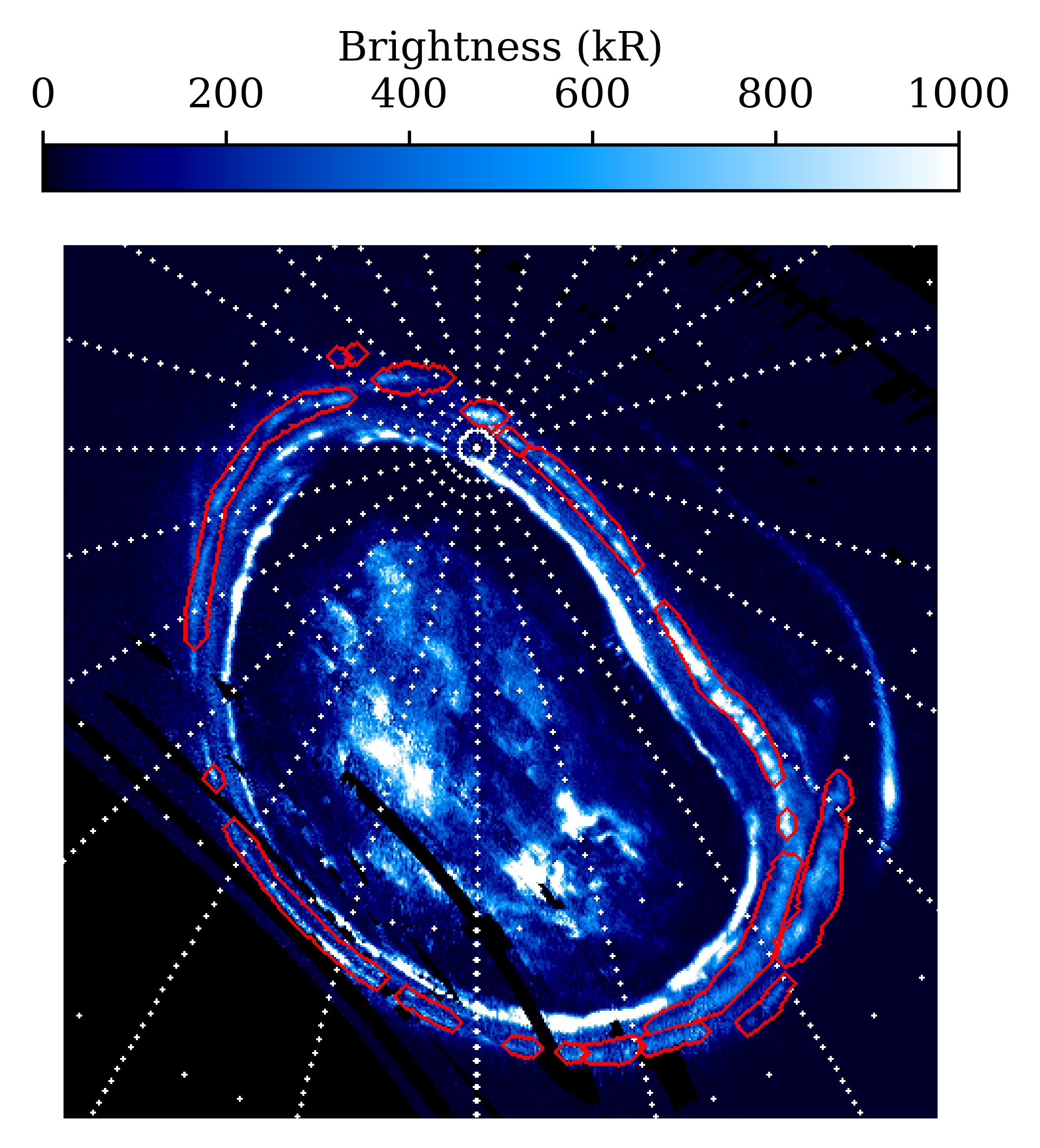}
    \caption{
        Same as Figure \ref{fig:rf_method_example_manual} but with automatically detected features.
    }
    \label{fig:rf_method_example_automatic}
\end{figure}

In order to determine the location of potential injection signatures in maps of the aurora, an algorithm was developed that combines manual feature designations with an automatic random-forest classifier.
This combines the accuracy of manual designation of which features constitute probable injection signatures with the objectivity of the automatic random-forest classifier.
A random-forest classifier is a classification model that uses a large number of decision trees to assign labels to data points \citep{ho:1998}.
Each decision tree is trained on a random subset of the data and assigns a label to each data point by taking the most common result from the ensemble of decision trees (a ``forest'').
The total number of decision trees in the forest and their length (number of branches/decisions in the tree) can be varied to produce labellings of varying accuracy to the original labels.
Random-forest classifiers can be applied to assign labels to unseen data in a partially labelled dataset or, as is the case in this work, to improve the objectivity of a set of manual labels.
By varying the number and length of the decision trees in the forest, the ``coarseness'' of the generated labelling can be modified, from unusably coarse for very small forests of very short trees up to a perfect recreation of the original labelling for very large forests of very tall trees.
Somewhere between these two extremes lies a middle ground that will return a labelling that is more objective than the original manual labelling while still approximating it in a broad sense.

Manual labelling of potential injection signatures was performed by designating polygons on maps of the aurora.
Since the colour-bar limits of these maps was set at (0, 250) kR, these manual designations are likely insensitive to injection signatures with maximum brightnesses below $\sim$50 kR, though this brightness is comparable with the background brightness of the diffuse aurora and thus injection signatures with this maximum brightness are, in any case, not likely to be easily differentiable from the rest of the aurora.
These polygons were designed to comfortably contain the entire injection signature rather than to follow its border exactly, since it was expected that further refinement of the injection-signature masks would be performed by the random-forest classifier.
Since these designations were performed ``by eye'', they are naturally more sensitive to injection signatures that are large in the ionosphere, and thus to regions of weak surface magnetic field.
This is not expected to significantly affect the results of this work as \textit{Juno} performs its perijoves without any bias in subsolar longitude, and hence a particular local time will correspond to many System-III longitudes during the first 40 perijoves. 

The random-forest classifier used in this work takes the pixel masks generated by the above manual-designation process and combines them with the pixelwise brightness, colour ratio, and projected radial distance, System-III longitude, and local time in the equatorial magnetosphere to produce automatic pixelwise masks of potential injection signatures.
Each data point to be assigned a label by the random-forest classifier is thus a polar-projected-image pixel represented as a vector with five properties.
The projected location in the magnetosphere is used alongside the (ionospheric) brightness and colour ratio to refine the behaviour of the detector; the brightness of injection features or the background aurora may vary based on local magnetic-field strength, local time, or distance from the main emission, and so it is important to encode these parameters for the detector.
It was a posteriori determined that the random-forest classifier does a very poor job of creating masks for injection signatures when only the brightness and colour ratio are used as inputs.
A standard 4:1 train-test split was used to train the classifier and assess its accuracy.
A 50-tree forest with tree depth of 100 was determined to adequately perform the classification.
Smaller forest sizes or shorter trees gave far less precise injection-signature masks, whereas increasing these parameters above these values, even by orders of magnitude, did not materially alter the performance of the classifier; the results of the 50-tree, 100-depth random-forest classifier represent a settled solution for this dataset.
The training of this classifier was performed using the \texttt{scikit-learn} Python package, and resulted in a model accuracy of 97\%.
Qualitatively, the masks produced by the automatic classifier strongly resemble the manual designations but smoother and more granular.
Some injection signatures are split or merged, some are absent, and extra injections have been identified, as expected from an automatic method, but it fundamentally identifies the same population of potential injection signatures as the manual designation. 
An example of manual designations of potential injection signatures and their automatically determined equivalents is given in Figures \ref{fig:rf_method_example_manual} and \ref{fig:rf_method_example_automatic}.
The trained classifier is available for download at \url{https://zenodo.org/records/17751841}.

\renewcommand{\thefigure}{B.\arabic{figure}}
\renewcommand{\thetable}{B.\arabic{table}}
\setcounter{figure}{0}
\setcounter{table}{0}
\setcounter{equation}{0}
\section{Sensitivity analysis for blob/arc longitudinal extent}
\label{sec:sensitivity_analysis}

\begin{figure}[tbhp]
    \centering
    \captionsetup[subfigure]{width=0.48\linewidth}
    \subfloat[Small blobs.]{
        \includegraphics[width=0.5\linewidth]{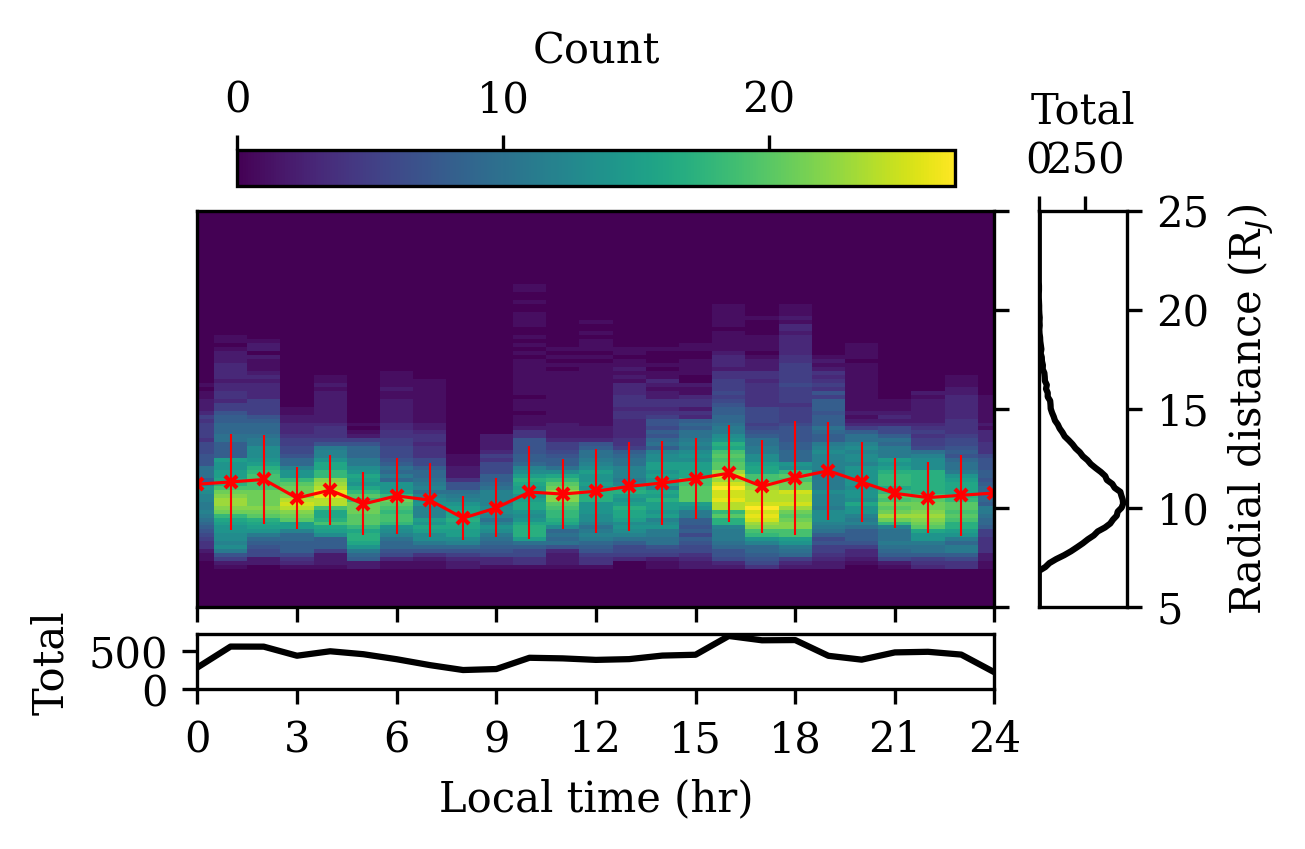}
    }\\
    \subfloat[Arcs.]{
        \includegraphics[width=0.5\linewidth]{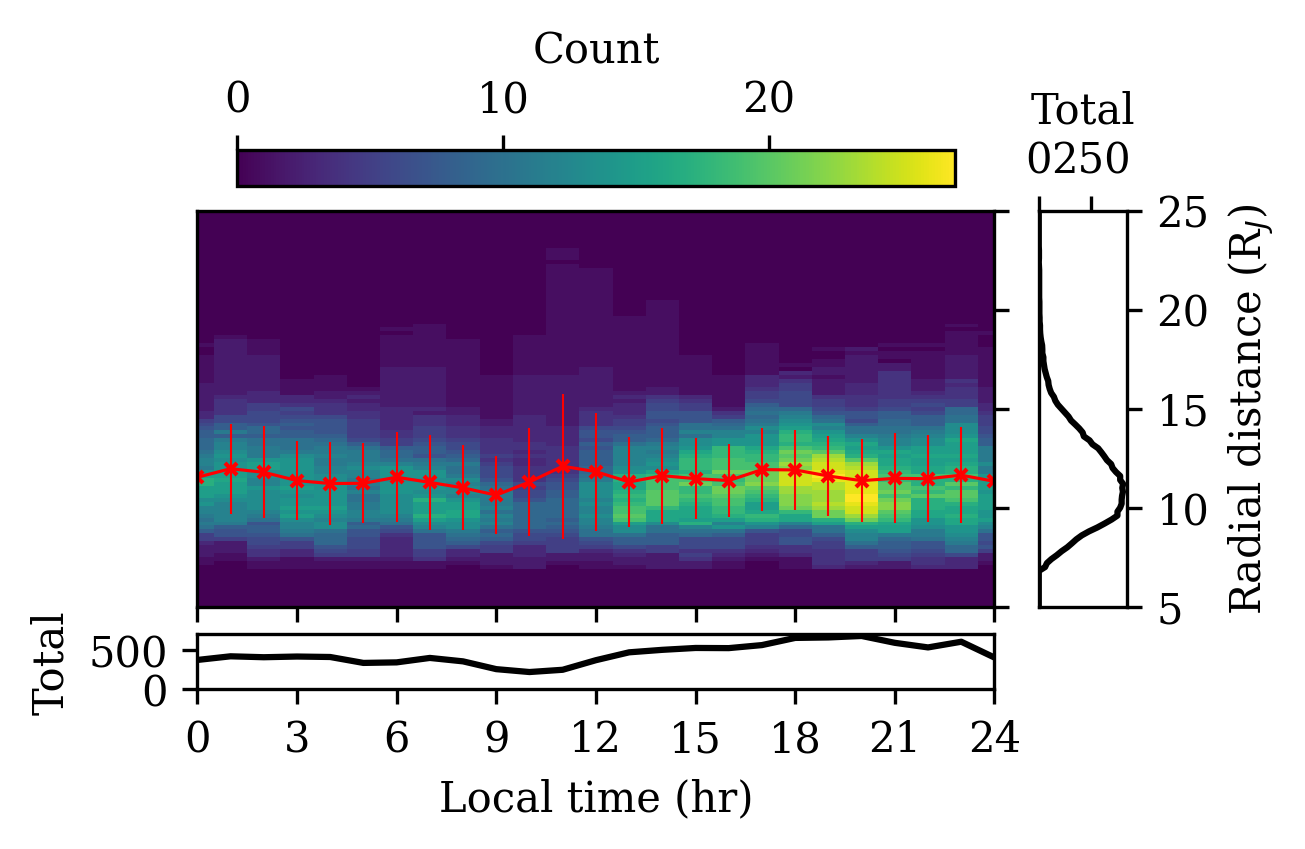}
    }
    \caption{
         Histogram in radial distance and local time of the projected position in the equatorial plasma sheet of features (longitudinal-extent cutoff = 20\textdegree) in the outer emission. The mean-average location for each local-time bin is given by the red line; error bars denote the standard deviation. Histograms flattened in local time and radial distance are given to the bottom and right of the main plot, respectively.  
    }
    \label{fig:location_histogram_low}
\end{figure}

\begin{figure}[tbhp]
    \centering
    \captionsetup[subfigure]{width=0.48\linewidth}
    \subfloat[Small blobs.]{
        \includegraphics[width=0.5\linewidth]{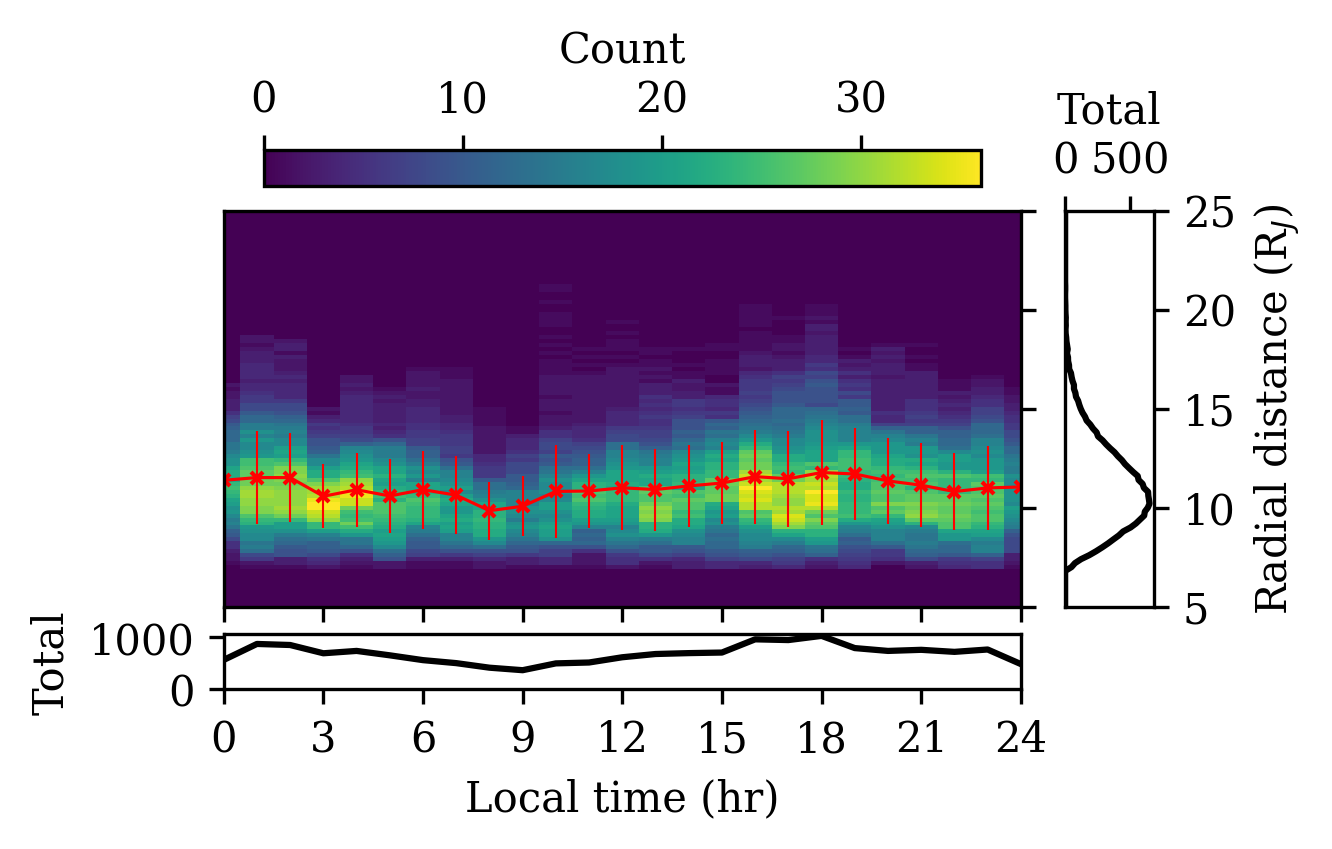}
    }\\
    \subfloat[Arcs.]{
        \includegraphics[width=0.5\linewidth]{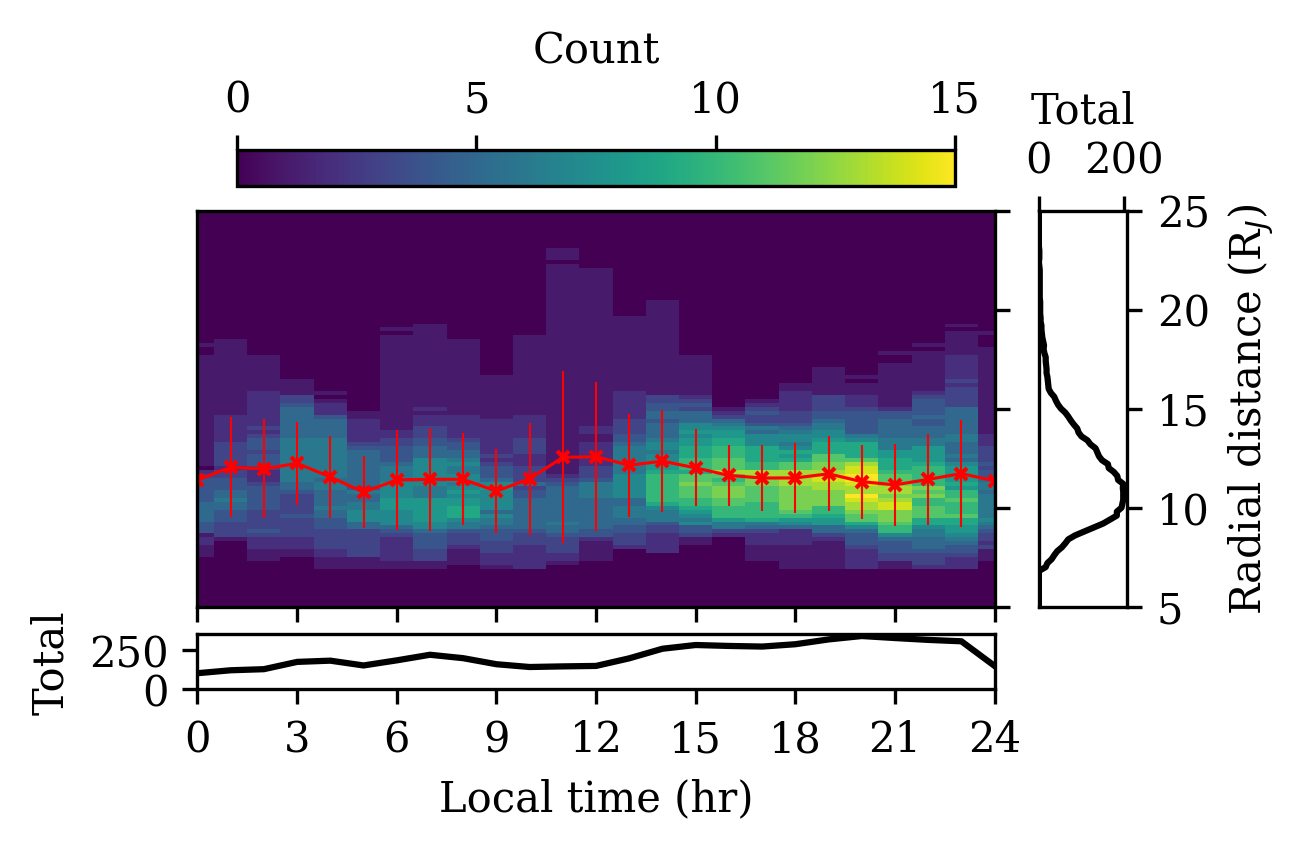}
    }
    \caption{
         As Figure \ref{fig:location_histogram_low} but with a longitudinal-extent cutoff of 40\textdegree. 
    }
    \label{fig:location_histogram_high}
\end{figure}

It can been seen in Figures \ref{fig:location_histogram_low} and \ref{fig:location_histogram_high} that varying the longitudinal-extent cutoff between automatically detected blob-like and arc-like features in the outer emission by $\pm$10\textdegree\ has little effect on the distribution shown in Figure \ref{fig:location_histogram} (longitudinal-extent cutoff = 30\textdegree).
In all cases, the small-blob features and arc-like features are both concentrated around a projected radial distance of $\sim$11 R$_{J}$, with arc-like features being more concentrated in the dusk sector than the noticeably more uniformly distributed small-blob features.
The choice of a 30\textdegree longitudinal-extent cutoff, while chosen as an approximate cutoff between the longitudinal spans of blob-like and arc-like features, is thus not expected to materially affect the conclusions of this work. 
Indeed, the fact that Figures \ref{fig:location_histogram_low} and \ref{fig:location_histogram_high} show such similar distributions is itself indicative that most blob-like features have longitudinal extents less than 20\textdegree, as assumed by \citet{dumont+:2014}, and that arc-like features mostly have longitudinal extents greater than 40\textdegree.

\renewcommand{\thefigure}{C.\arabic{figure}}
\renewcommand{\thetable}{C.\arabic{table}}
\setcounter{figure}{0}
\setcounter{table}{0}
\setcounter{equation}{0}
\section{Derivation of scattering surface-field-strength/power relation}
\label{sec:B_vs_power_derivation}

For the isotropic-scattering case, the precipitating electron flux is controlled by the loss-cone angle $\alpha$ in the ionosphere, which goes as
\begin{equation}
    \alpha=\sin^{-1}\left(\left(\frac{B_{MS}}{B_{IS}}\right)^{\frac{1}{2}}\right),
\end{equation}
where $B_{IS}$ is the magnetic-field strength in the ionosphere and $B_{MS}$ the magnetic-field strength at the source of the scattering in the magnetosphere \citep{mauk+:2017}, or 
\begin{equation}
    \alpha\simeq\left(\frac{B_{MS}}{B_{IS}}\right)^{\frac{1}{2}},
\end{equation}
since $B_{IS} \gg B_{MS}$.
If we assume, based on the area subtended by the loss cone for a given loss-cone angle, that the power $P\propto\alpha^{2}$, then
\begin{equation}
    P \propto \frac{B_{MS}}{B_{IS}}
\end{equation}
and hence the ratio of power in the northern and southern hemispheres, assuming that injection signatures are conjugate between hemispheres and hence arise from the same location in the magnetosphere, 
\begin{equation}
    \frac{P_{N}}{P_{S}} = \frac{B_{S}}{B_{N}}
\end{equation}
or equivalently
\begin{equation}
    \log_{2}\left(\frac{P_{N}}{P_{S}}\right) = -\log_{2}\left(\frac{B_{N}}{B_{S}}\right).
\end{equation}
This relation assumes that the variability of the intrinsic ``intensity'' or the average electron energy of the plasma injection has little effect on the ratio of hemispheric auroral power.
Different penetration depths (and hence different absorption profiles) between hemispheres may slightly affect this relation and hence account for some of the scatter in Figure \ref{fig:field_power_ratio}.

\renewcommand{\thefigure}{D.\arabic{figure}}
\renewcommand{\thetable}{D.\arabic{table}}
\setcounter{figure}{0}
\setcounter{table}{0}
\setcounter{equation}{0}
\section{Effect of sub-corotation on hemispheric power ratio}
\label{sec:subcorotation}

Figure \ref{fig:field_power_ratio} shows a large degree of scatter, which we suggest may be partially accounted for by the slight sub-corotation of injections.
The north-to-south auroral power ratio of injection signatures is typically greater than one for N$\rightarrow$S projections (i.e. where the injection signature is detected in the northern hemisphere and magnetically projected to the south) and less than one for S$\rightarrow$N projections.
In other words, the emitted power in the detected injection signature is larger than the emitted power in the projection of the injection signature in the other hemisphere.
This may be due to the slight sub-corotation of injections, which would mean that, for a given injection signature, the conjugate signature in the other hemisphere would be located slightly ahead or behind the original signature in magnetospheric System-III longitude.
To test this hypothesis, we consider a hypothetical injection signature with Gaussian power profile
\begin{equation}
    p = p_{0} \cdot \exp \left( -\frac{(\theta - \mu)^2}{2\sigma^2} \right), 
\end{equation}
where $p_{0}$ is the peak power, $\theta$ is the System-III longitude of a given point, $\mu$ is the central longitude, and $\sigma$ is the RMS width of the injection signature.
To calculate the total power $P$ of the injection signature, we integrate within a set of reasonable bounds for our detected signature, which we take to be $\mu \pm 2\sigma$, and so
\begin{equation}
    \begin{split}
    P &= p_{0} \int_{\mu - 2\sigma}^{\mu + 2\sigma} d\theta   \cdot \exp \left( -\frac{(\theta - \mu)^2}{2\sigma^2} \right) \\ &= p_{0} \cdot \left[\sqrt{\frac{\pi}{2}} \cdot \sigma \cdot \text{erf} \left( \frac{\mu - \theta}{\sqrt{2}\sigma} \right) \right]_{\mu - 2\sigma}^{\mu + 2\sigma}.
    \end{split}
\end{equation}
For the power $P_{\text{proj}}$ contained in the projected injection-signature bounds, we assume that the injection signature has moved by $\dot{\theta}t$, where $\dot{\theta}$ is the sub-corotation rate in degrees per hour and $t$ the \textit{Juno} hemisphere traversal time in hours, and hence
\begin{equation}
    P_{\text{proj}} = p_{0} \cdot \left[\sqrt{\frac{\pi}{2}} \cdot \sigma \cdot \text{erf} \left( \frac{\mu - \theta}{\sqrt{2}\sigma} \right) \right]_{\mu - 2\sigma + \dot{\theta}t}^{\mu + 2\sigma + \dot{\theta}t}.
\end{equation}
Since the Gaussian function is symmetrical about its mean, this expression is valid for both N$\rightarrow$S and S$\rightarrow$N projections, i.e. whether the conjugate injection signature is ahead or behind of its expected (fully corotational) projected location.
The ratio of these two powers is thus
\begin{equation}
    \frac{P}{P_{\text{proj}}} = \frac{\left[ \text{erf} \left( \frac{\mu - \theta}{\sqrt{2}\sigma} \right) \right]_{\mu - 2\sigma}^{\mu + 2\sigma}}{\left[ \text{erf} \left( \frac{\mu - \theta}{\sqrt{2}\sigma} \right) \right]_{\mu - 2\sigma + \dot{\theta}t}^{\mu + 2\sigma + \dot{\theta}t}}
\end{equation}
or equivalently
\begin{equation}
    \label{eq:p_proj}
    \frac{P}{P_{\text{proj}}} = \frac{2\cdot\text{erf}\left(\sqrt{2}\right)}{\text{erf}\left(\frac{\dot{\theta}t + 2\sigma}{\sqrt{2}\sigma} \right) - \text{erf}\left(\frac{\dot{\theta}t - 2\sigma}{\sqrt{2}\sigma} \right)}.
\end{equation}
The average vertical shift between the full linear relation in Figure \ref{fig:field_power_ratio} and the linear relation for the N$\rightarrow$S and S$\rightarrow$N cases separately is 0.57$\pm$0.1, or equivalently, since Figure \ref{fig:field_power_ratio} is given in log$_{2}$ space, 
\begin{equation}
    \frac{P}{P_{\text{proj}}} = 1.49\pm0.1
\end{equation}
and hence, by assuming that the \textit{Juno} hemisphere traversal time $t = 3\ \text{hr}$, that injection signatures are approximately of longitudinal width $\sigma = 10$\textdegree, and solving equation \ref{eq:p_proj} numerically, 
\begin{equation}
    \dot{\theta}=5.4 \pm 0.4\text{\textdegree} \text{hr}^{-1}
\end{equation}
which is equivalent to a corotation fraction of 85$\pm$1\%.
This is within the 80-to-90\% corotation fraction given by \citet{dumont+:2018} and hence we suggest that the slight sub-corotation of injection signatures combined with \textit{Juno}'s hour-scale traversal from the northern to the southern hemisphere may account for a significant proportion of the scatter present in Figure \ref{fig:field_power_ratio}.

\begin{figure}[tbhp]
    \centering
    \includegraphics[width=0.5\linewidth]{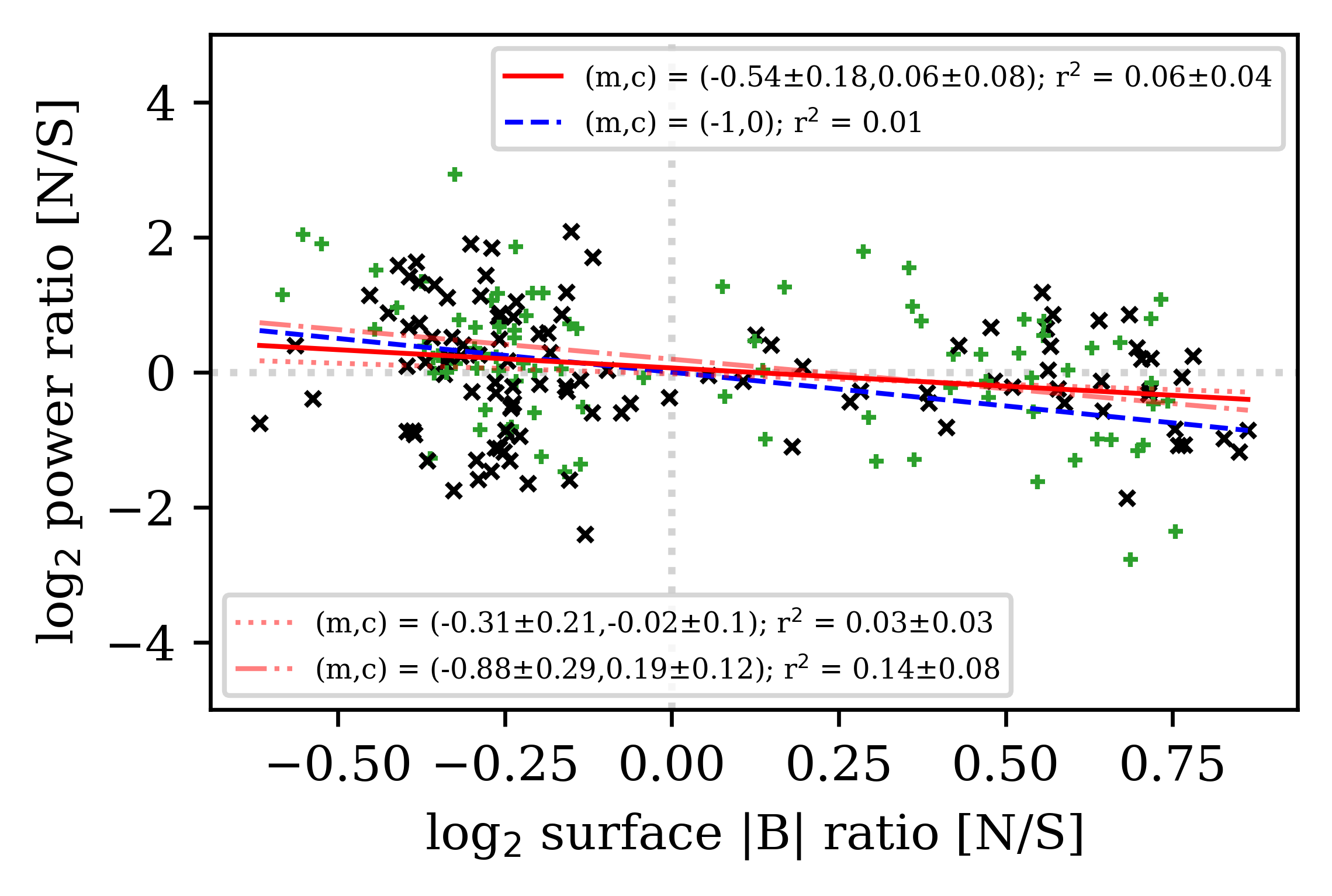}
    \caption{
        North-to-south UV auroral power ratio (conjugate area only) vs surface magnetic-field-magnitude ratio for injection signatures detected by \textit{Juno}-UVS during the first 40 perijoves. The N$\rightarrow$S projections are denoted by black $\times$, and the S$\rightarrow$N projections by green $+$. The best-fit linear relation for all points is given by a solid red line, and separate fitted relations for the N$\rightarrow$S and S$\rightarrow$N case by dotted and dash-dot lines respectively. The theoretical pitch-angle-scattering relation is given by a dashed blue line. 
    }
    \label{fig:field_power_ratio_corr}
\end{figure}

To test this hypothesis, a modified version of the analysis of Figure \ref{fig:field_power_ratio} was performed in which only those injections with a detected conjugate signature in the other hemisphere were used.
If the injection-signature polygon projected into the other hemisphere has less than 50\% of its area covered by a (presumed conjugate) detected injection signature, it is ignored.
Additionally, only the area of the two conjugate injection signatures that maps within the area of the other signature is considered when calculating the total emitted power ratio.
In principle, this should greatly reduce the effect of a slight sub-corotation on the emitted power ratio.
The results of this analysis are shown in Figure \ref{fig:field_power_ratio_corr}.
There remains some significant scatter in the data; however, the fitted linear relations for the N$\rightarrow$S and S$\rightarrow$N projections are no longer distinct from the overall linear relation.
In other words, the notion that the original detected injection signature always emits more power than its projected conjugate in the other hemisphere (present in Figure \ref{fig:field_power_ratio}) is not present in Figure \ref{fig:field_power_ratio_corr}.
We thus conclude that the slight sub-corotation of injections likely plays a significant role in explaining the distribution of points in Figure \ref{fig:field_power_ratio}, in line with the mathematical argument presented above.
It should be noted that injection signatures that are small enough to have no overlap in their detection polygons between hemispheres because of this sub-corotation are necessarily ignored by this analysis. 
This is expected to correspond to injection signatures less than 15\textdegree\ in longitudinal extent (since a feature moving at 85\% of corotation moves $\sim$15\textdegree\ in the $\sim$3 hours between \textit{Juno} passes of the two hemispheres), which may slightly affect the results in Figure \ref{fig:field_power_ratio_corr}.

\renewcommand{\thefigure}{E.\arabic{figure}}
\renewcommand{\thetable}{E.\arabic{table}}
\setcounter{figure}{0}
\setcounter{table}{0}
\setcounter{equation}{0}
\section{Supplementary figures}

\begin{figure}[tbhp]
    \centering
    \captionsetup[subfigure]{width=0.48\linewidth}
    \subfloat[17:41:35]{
        \includegraphics[width=0.5\linewidth]{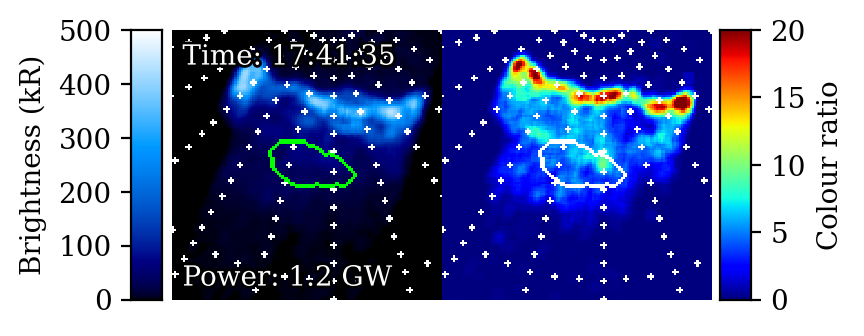}
    }\\
    \subfloat[18:44:44]{
        \includegraphics[width=0.5\linewidth]{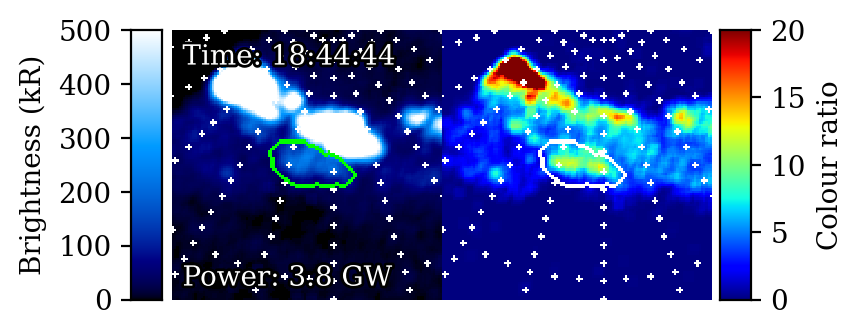}
    }
    \caption{
         An injection signature observed 2016-12-11 by \textit{Juno}-UVS during PJ3-S, highlighted in green. The brightness map is given on the left and the colour-ratio map on the right. The main emission is present above the injection signature. 
    }
    \label{fig:spontaneous_example_PJ3}
\end{figure}



\begin{figure}[tbhp]
    \centering
    \captionsetup[subfigure]{width=0.48\linewidth}
    \subfloat[02:52:48]{
        \includegraphics[width=0.5\linewidth]{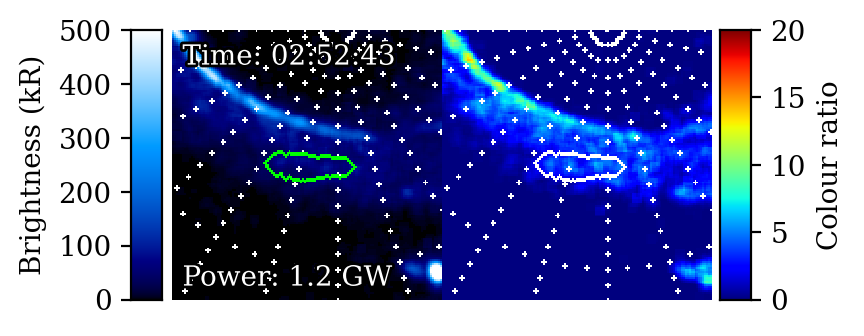}
    }\\
    \subfloat[04:21:10]{
        \includegraphics[width=0.5\linewidth]{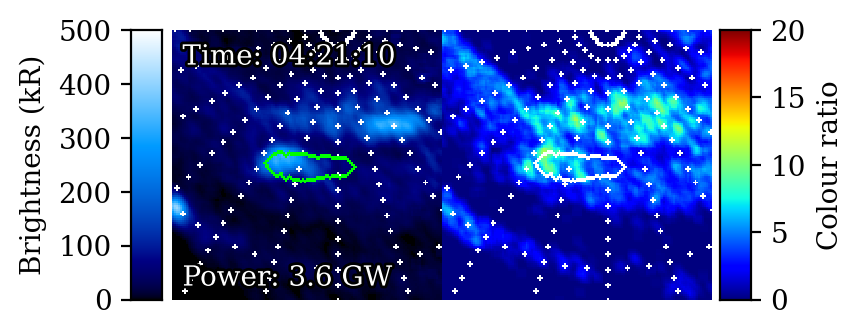}
    }
    \caption{
         An injection signature observed 2017-07-11 by \textit{Juno}-UVS during PJ7-S, highlighted in green. The brightness map is given on the left and the colour-ratio map on the right. The main emission is present above the injection signature. 
    }
    \label{fig:spontaneous_example_PJ7}
\end{figure}



\begin{figure}[tbhp]
    \centering
    \includegraphics[width=0.5\linewidth]{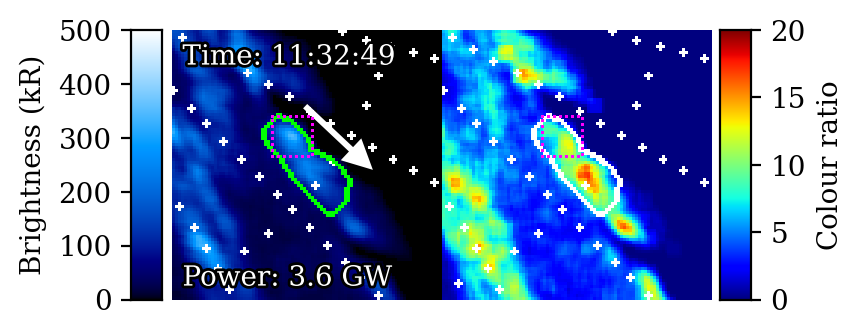}
    \caption{
        An injection signature observed at UTC 2019-04-06 11:32:49 by \textit{Juno}-UVS during PJ19-N, highlighted in green. The brightness map is given on the left and the colour-ratio map on the right. The main emission is present to the left of the injection signature. The direction of increasing magnetospheric System-III longitude is indicated by the white arrow. The position of the brightness peak is given by a dotted magenta square in both maps for the sake of comparison.
    }
    \label{fig:shift_example_PJ19_N}
\end{figure}

\begin{figure}[tbhp]
    \centering
    \includegraphics[width=0.5\linewidth]{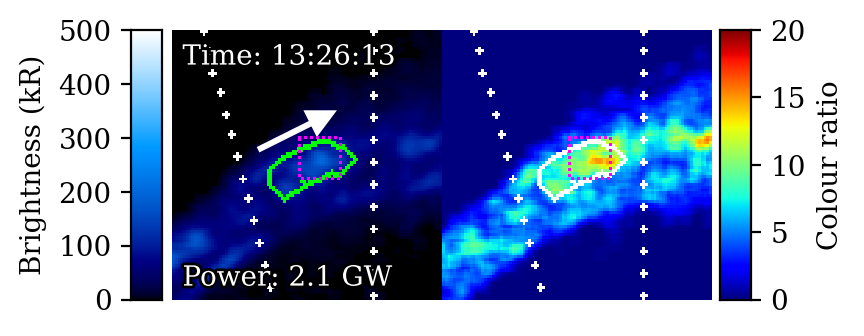}
    \caption{
        An injection signature observed at UTC 2019-04-06 13:26:13 by \textit{Juno}-UVS during PJ19-S, highlighted in green. The brightness map is given on the left and the colour-ratio map on the right. The main emission is present below the injection signature. The direction of increasing magnetospheric System-III longitude is indicated by the white arrow. The position of the brightness peak is given by a dotted magenta square in both maps for the sake of comparison.
    }
    \label{fig:shift_example_PJ19_S}
\end{figure}

\begin{figure}[tbhp]
    \centering
    \includegraphics[width=0.5\linewidth]{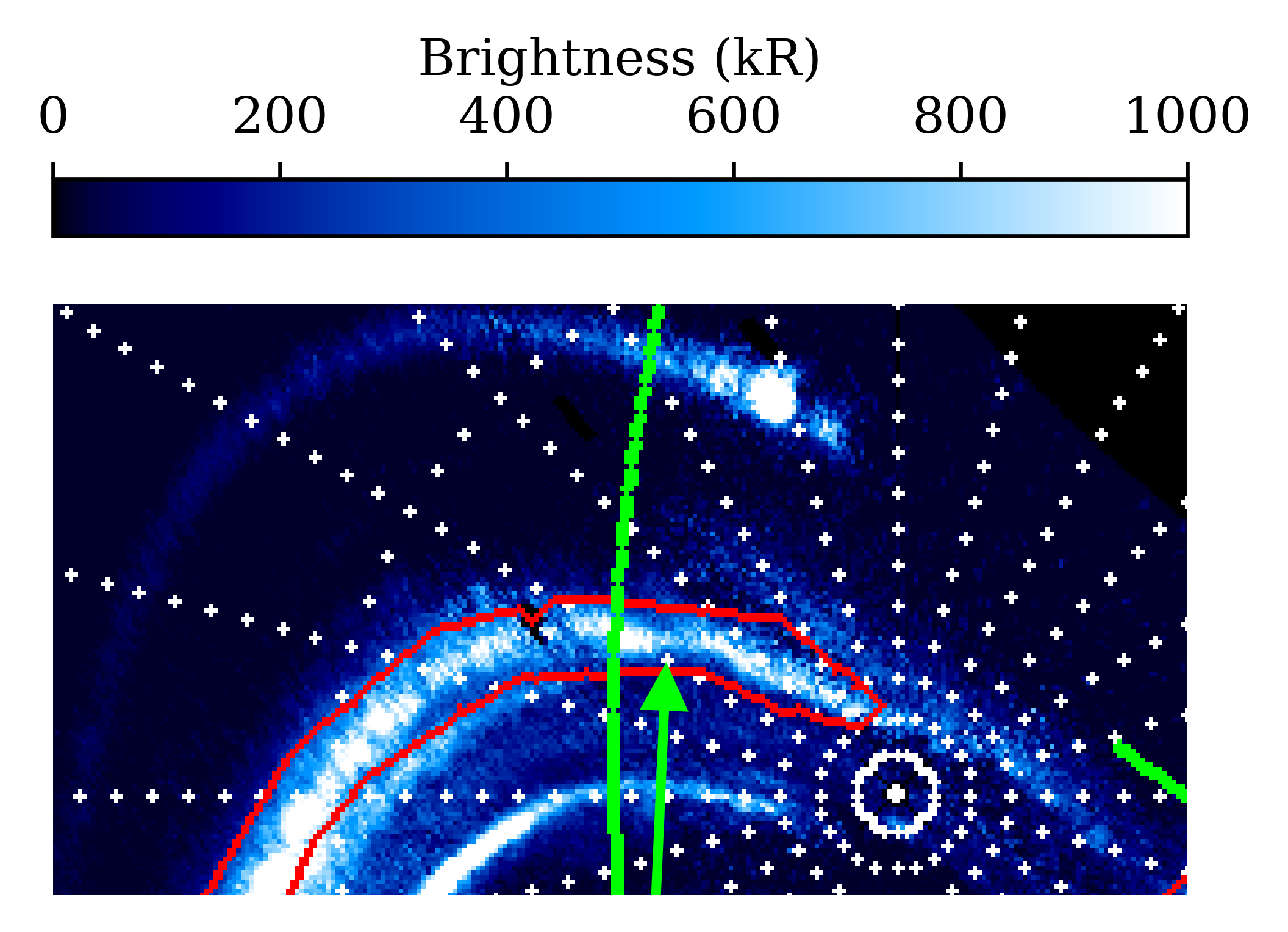}
    \caption{
        The \textit{Juno} footprint path (green) overlaid on the exemplar map of the aurora for PJ13-N. An automatically detected arc-type discrete feature crossed by \textit{Juno} is highlighted in red. The direction of travel of \textit{Juno} is denoted by the green arrow. 
    }
    \label{fig:crossing_PJ13_map}
\end{figure}

\begin{figure}[tbhp]
    \centering
    \includegraphics[width=0.5\linewidth]{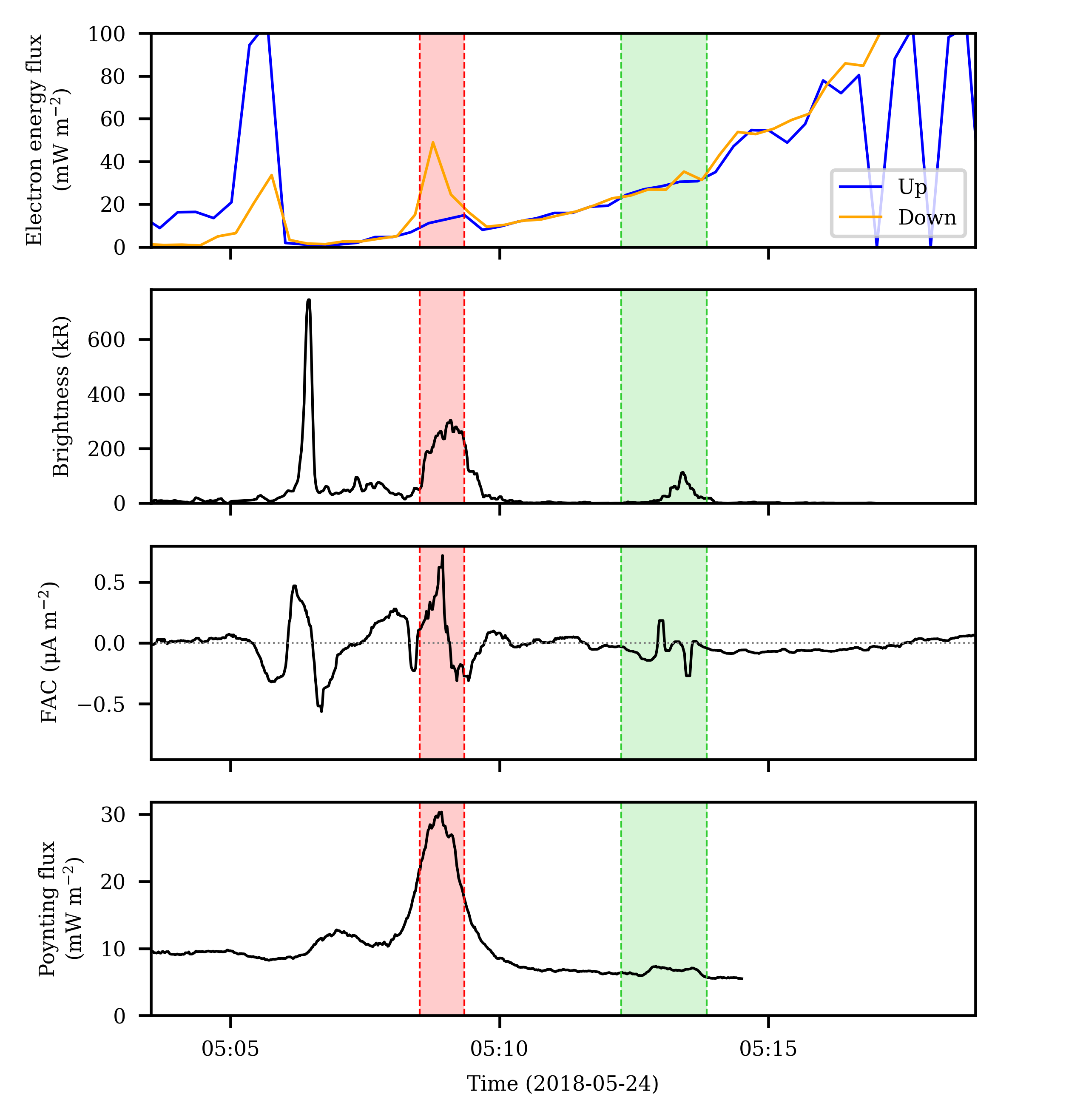}
    \caption{
        \textit{Juno} instrument data for PJ13-N. The crossing of the discrete feature in Figure \ref{fig:crossing_PJ13_map} is given in red, and the crossing of the Io footprint tail in green. From top to bottom: JEDI field-aligned (0\textdegree-20\textdegree, 160\textdegree-180\textdegree)  electron energy flux; UVS footprint brightness; calculated ionospheric field-aligned electrical current; calculated ionospheric Alfv\'{e}nic Poynting flux. 
    }
    \label{fig:crossing_PJ13_data}
\end{figure}

\end{appendix}

\label{LastPage}

\end{document}